\documentclass[12pt]{article}
\usepackage{geometry}
\usepackage{bbm}
\usepackage{verbatim}
\usepackage{amsmath}
\usepackage{amssymb}
\usepackage[authoryear]{natbib}
\usepackage[unicode=true,pdfusetitle,
 bookmarks=true,bookmarksnumbered=false,bookmarksopen=false,
 breaklinks=false,pdfborder={0 0 1},backref=section,colorlinks=false]{hyperref}
\usepackage{graphicx}
\usepackage{color}
\usepackage{caption}
\usepackage{multicol}
\usepackage{multirow}
\usepackage{amsfonts}
\usepackage{fancyhdr}
\usepackage[bottom]{footmisc}
\usepackage{version}
\usepackage{scalefnt}
\usepackage{subcaption}
\usepackage{rotating}
\usepackage{threeparttable}
\usepackage[export]{adjustbox}
\usepackage{bbm}
\usepackage{lscape}
\usepackage{threeparttable}
\usepackage[capposition=top]{floatrow}
\usepackage{comment}

\begin{document}
\title{\textbf{{
Procurement in welfare programs: Evidence and implications from WIC infant formula contracts\thanks{This paper supersedes an earlier version ``Who Pays for WIC's Formula? A Structural Analysis of
the US Infant Formula Market". We are grateful to Jorge Balat, Steve Puller, Aviv Nevo, Marc Rysman, Frank Verboven, Mo Xiao, Daniel Xu, Wei Zhao, and seminar participants at Central University of Finance and Economics, Fudan University, North Carolina State University, Vanderbilt University, Wuhan University, the 15th Annual International Industrial Organization Conference (2017), the 87th Annual Meeting of the Southern Economic Association (2017) for helpful comments or discussions. We thank the Economic Research Service of USDA and personnel from the Food and Nutrition Service of USDA for support collecting and organizing manufacturer bid data. }
}}}

\author{Yonghong An \and David Davis \and Yizao Liu \and Ruli Xiao\thanks{Yonghong An:
Department of Economics, Texas A\&M University, College Station, TX.
David Davis:
Ness School of Management and Economics, South Dakota State University, Brookings, SD. Yizao Liu:
Department of Agricultural Economics, Sociology and Education, Pennsylvania State University, University Park, PA. Ruli Xiao:
Department of Economics, Indiana University, Bloomington, IN; Email: rulixiao@iu.edu, phone:(812) 855-3213, fax: (812) 855-3736. }}
\date{\small{ July 2023}}
\maketitle
\vspace{-1.0cm}
\begin{abstract}
\footnotesize{
This paper examines the impact of government procurement in social welfare programs on consumers, manufacturers, and the government. We analyze the U.S. infant formula market, where over half of the total sales are purchased by the Women, Infants, and Children (WIC) program. The WIC program utilizes first-price auctions to solicit rebates from the three main formula manufacturers, with the winner exclusively serving all WIC consumers in the winning state. The manufacturers compete aggressively in providing rebates which account for around 85\% of the wholesale price. To rationalize and disentangle the factors contributing to this phenomenon, we model manufacturers' retail pricing competition by incorporating two unique features: price inelastic WIC consumers and government regulation on WIC brand prices. Our findings confirm three sizable benefits from winning the auction: a notable spill-over effect on non-WIC demand, a significant marginal cost reduction, and a higher retail price for the WIC brand due to the price inelasticity of WIC consumers. Our counterfactual analysis shows that procurement auctions affect manufacturers asymmetrically, with the smallest manufacturer harmed the most. More importantly, by switching from the current mechanism to a predetermined rebate procurement, the government can still contain the cost successfully, consumers' surplus is greatly improved, and the smallest manufacturer benefits from the switch, promoting market competition.  
}

\end{abstract}



\newpage

\section{Introduction}

Means-tested welfare programs, such as Medicaid, the Supplemental Nutrition Assistance Program, and the Temporary Assistance for Needy Families, are prominent in the U.S. On average, over 20 percent of all U.S. families participated in at least one major means-tested program per month, and the spending on federal programs expanded rapidly and reached \$1078 billion in 2020.\footnote{See \cite{foster2018program} and \cite{landers2021federal} for summary of federal welfare programs.} In many of those programs, the government procures products or services from private sectors using a variety of mechanisms and provides them to program participants.
Government procurement naturally separates consumers into two segments, one comprising non-participants and the other comprising less price-sensitive program participants. Market segmentation can distort manufacturers' behavior and affect non-participating consumers. The distortionary impacts depend on factors such as the procurement mechanism, the program size, the market structure, etc. 
Despite its importance in program evaluations and policy recommendations, there is very limited literature on government procurement's effects on markets and non-participating consumers. This paper aims to fill the gap by documenting such distortionary effects in the U.S. infant formula market, disentangling factors and channels driving the impacts, and simulating the effects of alternative procurement mechanisms.   

Over half of the sales in the U.S. infant formula market are procured for the WIC participants from the three dominating manufacturers -- Abbott, Mead Johnson (MJ), and Nestl\'{e}.\footnote{Nestl\'{e} purchased Gerber baby foods in 2007 and marketed infant formula under the Gerber Good Start brand name. Nestl\'{e}'s infant formula production and brands were purchased by Perrio in 2022. } 
In each state, the manufacturers compete to exclusively serve WIC participants through a first-price auction by bidding a rebate for their selected auction brands. The manufacturer with the lowest net price, computed as the wholesale price minus the rebate, wins a multi-year contract to serve exclusively WIC participants in the state. The auction data on rebates are striking: manufacturers submit extremely high rebates relative to their wholesale prices to win WIC contracts. Specifically, submitted rebates exceed 80\% and 90\% of wholesale prices for more than 75\% and 25\% bids, respectively, which significantly contain the cost of the program. 

Motivated by the success of the auction, we answer five crucial questions in this market and beyond. First, what are the underlying determinants of the high rebates offered by the manufacturers? Second, what role does the WIC program size play in the impact of the auction procurement? Third, how does the connection between the two market segments affect non-WIC consumers' retail prices? Would it incentivize the manufacturers to charge a higher retail price so that the non-WIC consumers partially bear the rebate provided to the government? Fourth, how does the procurement auction affect the market structure, particularly the competition among the manufacturers? This question is of particular importance because Wyeth-Ayerst (Wyeth), the fourth major infant formula manufacturer, exited the market in 1996, several years after the onset of the auction mechanism. Competitive bidding for the exclusive right to serve WIC participants is one of the main driving forces of the exit.\footnote{Moore, A. K. (1996, January 29), Wyeth-Ayerst Leaving U.S. Formula Market. Retrieved from https://www.supermarketnews.com/archive/wyeth-ayerst-leaving-us-formula-market.} Lastly, would the other prevalent procurement mechanism - a predetermined rebate chosen by the government - be better for consumers, manufacturers, and the government? The answer to this question is vital for policymakers to design future procurement rules. We answer all the aforementioned questions by presenting reduced-form evidence, proposing and estimating a structural model of the market, and conducting counterfactual analyses. 

We first conduct reduced-form analyses to document the significant impact of winning a WIC contract on the non-WIC markets. We find that non-WIC sales rise significantly upon winning a WIC contract.\footnote{\cite{huang2014wic} demonstrate that a spillover effect is likely essential for manufacturers to compete in the market.} Compared with losing an auction, winning boosts the winner's non-WIC sales for auctioned and non-auctioned brands by about 80\% and 130\%, respectively. Retail prices also rise upon winning. Specifically, the retail price of the winner's auction brand is 8\% higher than that from losing. The prices of its non-auction brands rise with a smaller magnitude but are not statistically significant. These results indicate that the WIC auction procurement might distort the non-WIC market outcome despite the price regulation by the federal government. 

To explore and disentangle the determinants and channels of the distortionary effects, we propose a structural model to characterize the behaviors of manufacturers and non-WIC consumers in the infant formula market. We model the demand of non-WIC consumers in the traditional random coefficient discrete choice fashion and allow non-WIC consumers to prefer the products of the winning manufacturer. On the supply side, manufacturers compete sequentially in two stages: they submit rebates to compete for the WIC contract in the first stage and then set retail prices after observing the auction outcome. We mainly focus on the pricing competition in the second stage and rely on a flexible reduced-form analysis to capture the determinants of the rebates in the auction stage.

Several new features distinguish our model of price competition from the existing ones. First, the auction winner faces two demand segments - regular non-WIC consumers and price-insensitive WIC consumers. Second, the winning brand's retail price is regulated to have an implicit cap, and the two segments cannot be discriminated in price. The existence of price-insensitive consumers and the price regulation in the market impose unprecedented challenges on our analysis of the retail pricing game. On the one hand, economic theory suggests that if consumers are price-insensitive, the price should be as high as possible. On the other hand, price regulation effectively restricts the equilibrium price that can be reached. The two features work in opposite directions, resulting in a slightly higher equilibrium price from our reduced-form analyses. It is unclear in the existing literature  how to model price competition incorporating both features. To tackle this challenge, we propose a novel sequential pricing model, where, first, the winning manufacturer determines the price for the WIC brand based on the non-WIC demand, then raises the equilibrium price for the brand for both segments of consumers based on the information of demand and regulations. We allow such a pricing strategy to be manufacturer-specific. Second, given the price of WIC brands, the manufacturers conduct Bertrand price competition to determine the prices of other brands. It is worth noting that we also allow possibly lowered marginal costs for winning manufacturers, capturing cost savings from the economies of scale, less promotion and advertisement spending, etc.  

We estimate the model using sales data from NielsenIQ and manufacturers' rebates data from the U.S. Department of Agriculture (USDA)'s Food and
Nutrition Service (FNS). The first step is to estimate the consumers' preferences using the BLP approach (\cite{berry1995automobile}), confirming a sizeable spill-over effect of winning a WIC auction on the demand of non-WIC consumers. This indicates a great advantage for winning manufacturers in the non-WIC market. Next, we recover the marginal cost using the estimated demand parameters and the observed  retail prices. We find considerable cost savings upon winning, and all three manufacturers raise the prices of their WIC brands after winning. These findings rationalize the significant rebates provided by manufacturers and address the first question raised. 

We further show that, on average, winning manufacturers' after-rebate prices to WIC agencies are significantly lower than their corresponding marginal costs for more than half of the winning markets, resulting in negative profits in the WIC market. Nevertheless, their loss in the WIC market is compensated by extra profit in the non-WIC market, in which the winners enjoy a high price-cost margin (around 50\%) due to the substantial spillover effect and considerable cost savings. We also found that the marginal cost is significantly lower than the wholesale price, suggesting that manufacturers enjoy large market power in the non-WIC market. 

Using the demand and supply estimation results, we answer the four remaining questions by simulating the market outcomes, including manufacturers' profits, consumer surplus for both WIC and non-WIC consumers, and government expenditures in different hypothesized market mechanisms. First, to fully evaluate the impact of the WIC auction, we simulate and compare the market outcomes with and without the WIC auction, where in the latter scenario, the government provides a voucher for quantities of food, as in the case of all food products provided by the WIC program. We find that (1) without the WIC auction, consumer surplus would increase 11\%, with WIC and non-WIC consumers better off and worse off, respectively. WIC consumers are better off because of the availability of more options\footnote{Under the current WIC program, infants can only choose the brand contracted by the state agency, which restricts the WIC participants' choices and, thus, their welfare is affected negatively. The consequences of such restriction are even more severe during the 2022 infant formula shortage.}. Surprisingly, non-WIC consumers would be worse off. This is due to the lower retail prices resulting from 
the spillover effect and cost savings enjoyed by the winner when the WIC auction is used for procurement. (2) The WIC auction significantly reduces manufacturers' profits. Nestl\'{e}, the smallest manufacturer, faces the most significant negative impact of the auction. Compared with procured without auctions, the profit of Abbott, MJ, and Nestl\'{e} drops by 42\%, 58\%, and 75\%. (3) As expected, the government can contain the cost using the auction mechanism. Specifically, the government's monthly expenditure reduces by 65\% due to the WIC auction. 

We next investigate how WIC program size affects the market outcomes, mainly the manufacturers' profits in the current mechanism with the auction. We find that the larger the program size, the higher profits for the two larger manufacturers. In contrast, the profit of the smallest manufacturer Nestl\'{e} decreases in the WIC program size. Our finding confirms the statement of Wyeth, which had a similar market share as Nestl\'{e}, that ``$\cdots$ the increasing growth of the WIC Program contributed to the decision" of exiting the domestic market. An important implication of our finding is that the impacts of WIC auctions on manufacturers are asymmetric and small ones suffer more. This might imply that the WIC auction is an entry barrier in the infant formula market. 

Finally, we explore another prominent procurement mechanism: predetermined rebates, employed by the Medicaid drug rebate program (MDRP). Under such a mechanism, manufacturers voluntarily join the program and provide products or services to program participants by accepting the rebates determined by the government \textit{ex ante}. We simulate the market outcomes if the government switches from competitive bidding procurement to predetermined rebates. We find that in the mechanism of predetermined rebates, the government pays 27\% more  relative to the competitive bidding. However, both consumer surplus and total surplus would be higher. The surplus of WIC consumers increases substantially (25.5\%) while that of non-WIC consumers decreases slightly (5.1\%). Manufacturers, again, are affected asymmetrically. Nestl\'{e}'s profit would be 47.3\% higher while that for Abbott and MJ would be 30.3\% and 14.8\% lower, respectively.  

This paper contributes to the literature by thoroughly analyzing the infant formula market, explaining the unreasonably high rebate, and evaluating the impacts of WIC on nonparticipants. To the best of our knowledge, this is the first work to model and estimate pricing competition with two market segments, where one is price inelastic.\footnote{\cite{abito2022demand} also study the WIC program. Their focus is on the spill-over effect in the demand size. They assume manufacturers conduct Bertrand pricing competition on the supply side but only maximize non-WIC profits without incorporating the two market segment features. Moreover, they do not model the price regulation by the government, which results in a negative marginal cost estimate.} 
Using an event study, \cite{oliveira2004wic} argue that non-WIC consumers are price insensitive. Thus, retailers increase the contracted brand's price. Similarly, using a Cournot oligopoly model, \cite{prell2004economic} also attributes the substantial rebate and relatively high wholesale prices to the demand elasticities of consumers. Our paper is fundamentally different from the existing work in that (1) we model manufacturers' pricing strategies in non-WIC and WIC markets jointly by a sequential game; (2) we utilize both sales data and manufacturers' rebates for estimation; and (3) we rigorously investigate the impacts of the WIC program by counterfactual analyses.

Our paper is also related to other studies on infant formula, focusing on different topics. \cite{huang2014wic} and \cite{oliveira2010rising} identify a spillover effect as an essential factor in setting prices for infant formula. \cite{davis2011bidding} considers the spillover effect and shows that manufacturers' marginal costs are often lower than the after-rebate prices. Hence, the WIC program has no impact on non-WIC consumers. \cite{black2004special} and \cite{miller1985impact} document the effects of WIC formula on the health outcomes of infants. More generally,   \cite{meckel2020cure,li2022impacts} study impacts of Electronic Benefit Transfer on the WIC program. \cite{mclaughlin2019economics} focus on authorized vendors that derive more than 50\% of their food sales through WIC. 

Our counterfactual analyses shed light on how government procurement may affect market competition and nonparticipants' behavior toward other welfare programs. Although the infant formula market is uniquely characterized by the rebate program and high concentration, our analysis and empirical results can provide useful tools and policy implications to other markets where government programs affect related markets. The existing literature on welfare programs is largely silent on the programs' distortionary effects on the market and nonparticipants. A notable exception is \cite{duggan2006distortionary}, which finds that government procurement methods in the Medicaid program can change the equilibrium price of pharmaceuticals and product proliferation in the private sector. Our paper fills the gap by providing empirical evidence that WIC's rebate and auction procurement method serves as a barrier for manufacturers with small market shares to enter or grow in the market.

Finally, this paper is also related to the literature on consumer subsidy policies where manufacturers compete for their products' eligibility for a subsidy. \cite{fan2022welfare} studies a Chinese cellphone subsidy program ``Home Appliances Going to the Countryside," and finds that such competition mitigates price increases and improves consumer and total surpluses while limiting government subsidy payments. 
Our paper is different in that we explicitly model manufacturers' competition for eligibility, while \cite{fan2022welfare} assumes the eligibility is given exogenously.

The outline of the paper is as follows. Section 2 provides institutional background on the US infant formula market and describes the data used for analyses. Section 3 presents some reduced-form analyses of the effects of winning a WIC contract on sales and prices. Section 4 proposes a structural model of manufacturers' two-channel competition. Section 5 describes the identification and estimation strategies. Section 6 presents the estimation results.  Section 7 conducts several counterfactual analyses, and Section 8 concludes. Figures and tables are in the appendix.

\section{The Institutional Background and Data \label{section: market and data}}

\subsection{The WIC program}
The WIC program, established in 1972, provides various services and supplemental foods for low-income women, infants, and young children under five years of age. The program is administered jointly by the FNS of USDA and authorized state agencies. A family member at nutritional risk is eligible if (1) the family income is less than 185\% of the U.S. poverty income guidelines or (2) the family is enrolled in the federal Medicaid, Food Stamp, or Temporary Assistance for Needy Families  programs, even if the family income exceeds 185\% of the poverty line.\footnote{States may use enrollment status in other means-tested programs, such as the National School Lunch Program or the Supplemental Security Income program, to qualify an applicant as automatically eligible for WIC.} The program participants receive vouchers to redeem at authorized retail stores. State agencies reimburse retailers for the items sold to WIC participants based on the redeemed vouchers. WIC provides infant formulas for participated infants from 0 to 12 months. During our data period 2006-2015, 47.4\% infants participated in the WIC program. 

WIC state agencies purchase infant formulas using a competitive bidding process (sealed-bid auction) under the federal law implemented in 1989 to contain the procurement costs. WIC state agencies can solicit bids independently or as part of multistate alliances.\footnote{Currently, there are five multistate alliances and territories and Indian tribal organizations. Mountain Plains (Missouri, Nebraska, South Dakota); Southwest, Mountain Plains, and Midwest (Iowa, Minnesota, and Texas); New England and Tribal Organizations (Connecticut, Maine, Massachusetts, New Hampshire, and Rhode Island); Western States Contracting Alliance (Alaska, Arizona, Delaware, District of Columbia, Hawaii, Idaho, Kansas, Maryland, Montana, Nevada, Oregon, Utah, Washington, West Virginia, and Wyoming); Southwest Region (Oklahoma); and Southwest/Southeast (Arkansas, New Mexico, and North Carolina). In 2004, Congress limited the size of new alliances to 100,000 participating infants and limited the ability of current alliances with over 100,000 participating infants to expand.} Those states with home delivery (Vermont) or direct distribution (Mississippi) or Indian State agencies with 1,000 or fewer participants are exceptional and excluded from our analysis. The auction proceeds as follows. Infant formula manufacturers submit sealed bids that specify a rebate for the brands they choose to participate in the program, i.e., auctioned brands. Those auctioned brands usually are infant formulas suitable for routine issuance to generally healthy, full-term infants. The contract is awarded to the manufacturer asking for the lowest net price, calculated as wholesale price subtracts rebate. A contract typically lasts three years; occasionally, it can be extended via negotiation between the winning manufacturer and the state agency. WIC infants can only get the contracted brand during the contract period.\footnote{Most state agencies also provide alternative therapeutic formulas for those infants with special needs if a medical documentation form or medical prescription is provided. The percentage of therapeutic formulas is negligible. } 

The cost of infant formula to the WIC program is the difference between the WIC brand retail price and the winning manufacturers' rebate. The state agency regulates the retail prices of WIC infant formula. The motivation for such regulations is that WIC consumers are price insensitive, so profit optimization manufacturers/retailers might charge a very high retail price, which would work against the cost containment objective. Even though the regulation details vary across states, the main limitations on retail prices are that they must be competitive with, and within the limitations for, similarly situated vendors, e.g., in terms of type and size of the store and geographic location. Such a price restriction is imposed as a requirement for obtaining the eligibility to carry WIC products in the state.\footnote{In addition to price regulations, there are some other selection criteria for authorizing retailers to carry WIC products, e.g., a minimum stock of WIC products, accessibility to participants and WIC staff, business integrity, etc.} Only authorized retailers are eligible to obtain reimbursement from the redeemed vouchers. An example of the regulation is that the WIC agency sets price ceilings, called maximum allowable redemption rates or maximum allowable reimbursement level (MARL), by peer group for each food instrument the group redeems. Different states impose different regulations to implement the MARL. Some states set a specific number as MARL, e.g., Pennsylvania. Others set it to be a percentage of prices for the peer group, e.g., Texas.

\subsection{The infant formula market and data}
\subsubsection{The market}
Infant formula is vital for formula-feeding babies' healthy growth and development in their first year. In the U.S., approximately 54\% and 74\% of infants in 2018 were partially or fully fed with formula after 3 and 6 months, respectively.\footnote{https://www.cdc.gov/breastfeeding/data/nis\_data/results.html, retrieved May 27, 2022.} The U.S. infant formula market size is \$3.65 billion in 2019 and is projected to increase by 5.8\% annually. There are three major manufacturers in the market: Abbott, MJ, and Nestl\'{e} with their leading infant formula brands being Similac, Enfamil, and Gerber, respectively.\footnote{Wyeth was active in the domestic market until 1996 and discontinued its premium formulas in the US. In 1997, Wyeth reentered the US market as a producer of infant formula for PBM Products, Inc., which produces most of the generic-branded infant formula in the US and holds about a 1\% market share. The generic formula is usually sold at much lower prices compared to the products of the three major manufacturers.} The three manufacturers dominate the market, with their domestic market shares being 39.9\% for Abbott, 39.7\% for MJ, and 15.2\% for Nestl\'{e} in 2005-2016. Each manufacturer produces its infant formula in several production centers. The production centers of Abbott are in Casa Grande, AZ; Columbus, OH; Sturgis, MI; and Alta Vista, VA. MJ's are in Evansville, IN; Zealand, MI; and Springfield, MO. Nestl\'{e} only has one production center in Eau Clair, WI.

The main products in the market are milk- and soy-based infant formula with three physical forms: liquid concentrate, powder, and ready-to-feed. Milk-based powder dominates the market with a 78.3\% of market share, while the shares of milk-based liquid concentrate, milk-based ready-to-feed, soy-based powder, liquid concentrate, and ready-to-feed are 4.4\%, 7.0\%, 8.6\%, 1.0\%, and 0.7\%, respectively. To simplify our analysis, in this paper, we focus on milk-based powder, which involves 20 brands from the three leading manufacturers. We also exclude formula brands that target children between 12-24 months because WIC provides infant formulas for participating infants from 0 to 12 months only.

\subsubsection{Data}
Our data are from a variety of sources: (1) the price, volume, and sales of infant formula are from the Retail Scanner data from NielsenIQ; (2) nutritional characteristics of infant formula are collected from brands' websites; (3) the national wholesale prices and bids for WIC auctions are from the Food and Nutrition Service of USDA;  (3) demographics and cost shifters of infant formula are from various  sources. 

The Retail Scanner data provide information from 2006 to 2015 on infant formula's dollar and volume sales in supermarkets, grocery stores, and drug stores from more than 90 participating retail chains in 48 states (no data on sales in Alaska and Hawaii) and the District of Columbia. In addition, the data consist of detailed information on product characteristics, e.g., brand names, container sizes, package sizes, etc., and location and time of sales.  

The dollar sales and volume of infant formula for non-WIC consumers are calculated using the Retail Scanner data, the number of WIC and non-WIC infants, as well as the breastfeeding rates of WIC and non-WIC infants at the state-year level. The main idea is based on the representative nature of the NielsenIQ sales data, which allows us to reasonably approximate the ratio of non-WIC formula sales to the total sales in our data by using the proportion of formula-feed non-WIC infants to all formula-feed infants in the general population. The construction details are in Appendix \ref{Appendix: Data}.

\begin{figure}[tbp]
\caption{Average price and share by manufacturer across years}
\label{fig:price and share across years}\centering
\begin{subfigure}[t]{0.5\textwidth}
        \includegraphics[height=2.6in]{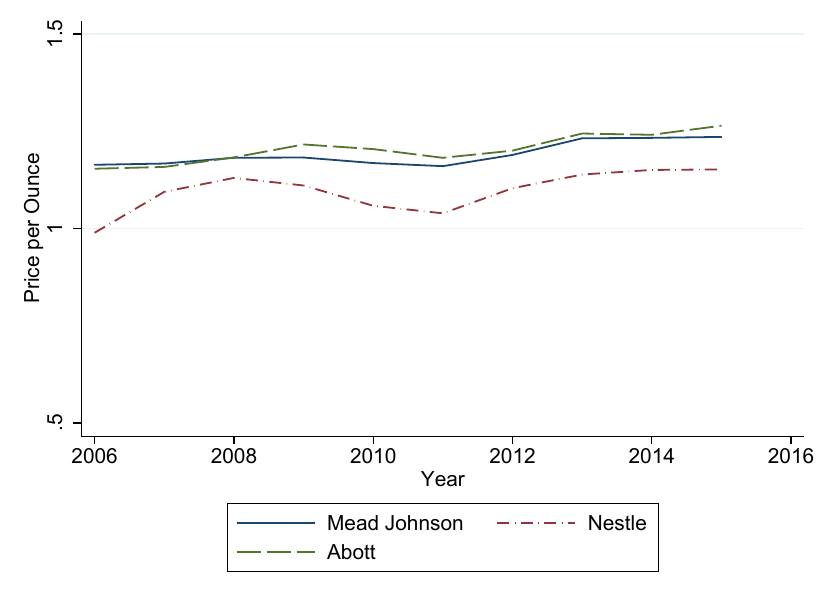}
    \end{subfigure}
\begin{subfigure}[t]{0.49\textwidth}
        \includegraphics[height=2.6in]{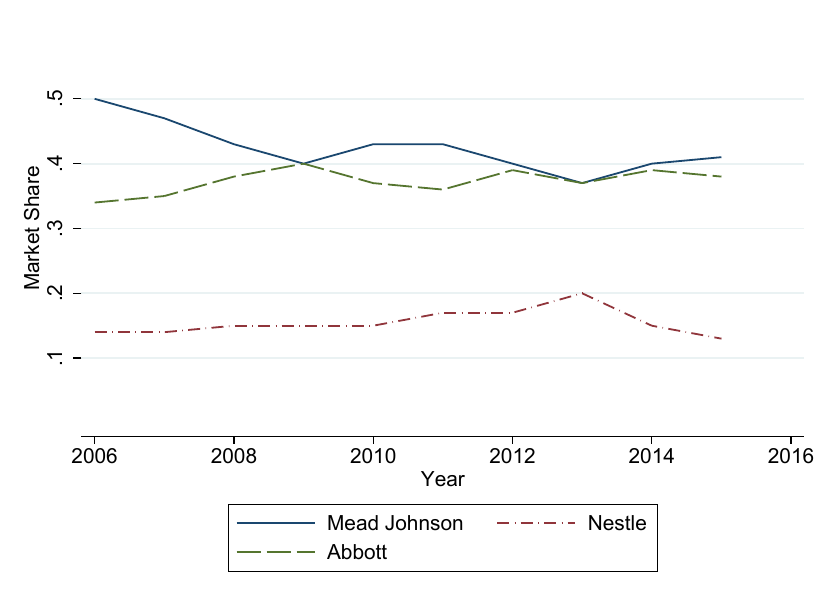}
    \end{subfigure}
\par
\vspace{0.3cm}
\begin{minipage}{0.7\textwidth} 
{\footnotesize \textit{Note:} Milk-based powder formula only. 
\par}
\end{minipage}
\end{figure}

We present summary statistics for milk-based powder across all states and years in Table \ref{tab:SummaryStatistics}. At the state level,  the average monthly non-WIC dollar sales for all the brands is \$444,430, the average monthly volume sold is 391,730 ounces, and the average sales-weighted retail price is \$1.16 per oz. MJ leads the market with an average monthly sales of \$626,580, followed by Abbott and Nestl\'{e}. The prices for MJ and Abbott are \$1.21 and \$1.22 per oz, respectively, which are very close. Nestl\'{e} charges a much lower price \$0.94 per oz. We display the average prices and market shares for the three manufacturers from 2006-2015 in Figure \ref{fig:price and share across years}, from which the market structure is very robust for our sample period: Nestl\'{e} continues to be the smallest manufacturer and to ask for a significantly lower price among the three dominant manufacturers  

The auctioned brands of milk-based powdered formula for the three manufacturers are Similac Advance, Enfamil Infant, and Gerber Good Start. In most cases, only one brand, known as the auctioned brand, is provided to all WIC infants from the manufacturer who emerges as the winner in the WIC auction. For instance, in Pennsylvania in 2016, Similac Advance from Abbott was the exclusive milk-based formula for WIC infants. There are occasional instances in certain states where more than one brand from the winning manufacturers is supplied to WIC infants. For example, in Connecticut, there are occasions when Similac Sensitive is provided alongside Similac Advance, which is the contracted brand. Nonetheless, the sales of these additional brands are generally minimal. To simplify our analysis, when multiple WIC formulas are available in a state, we aggregate all the products into a single auctioned brand. Subsequently, we calculate the sales-weighted nutritional characteristics for the aggregated auctioned brand. 
In this study, we focused on two specific characteristics of infant formula: anti-spit-up and probiotics. While there are other essential characteristics in infant formula, such as iron or DHA, we were unable to include them because almost all the formula products contain iron and DHA. 

The WIC auction data include a panel of the three manufacturers' annual national wholesale prices and rebates for milk and soy-based infant formula with three physical forms in 48 states and the District of Columbia (DC) from 1988 through 2015.\footnote{ Vermont used a home-delivery system, and Mississippi used a direct distribution system during our study period. Therefore, we exclude them from our analysis. Both states have now transitioned to retail delivery with the adoption of EBT.}  The details of the WIC contracts, e.g., previous winner, starting and ending time, and alliance status, are also available in the data. Wyeth also participated in the WIC auctions from 1988 to 1996 but exited the market afterward. To avoid modeling Wyeth leaving the market, we use the auction data from 1998, after which only MJ, Abbott, and Nestl\'{e} are potential bidders.

The WIC auctions can be categorized into uncoupled and coupled ones. Specifically, in the coupled auctions, milk and soy-based formulas either have the same percentage rates or their rebates are determined by complicated and unknown procedures. In the uncoupled auctions, the state agency hosts separately one auction for milk and one for soy formula. We only keep auctions for milk-based powder, including the uncoupled and those coupled ones with milk and soy-based formula having the same percentage of rebates.\footnote{In the data, the winner of a contract is always consistent with the winner of the milk-based power auctions.} 

Table \ref{table: sumstat of whole sale and rebate} presents summary statistics of the three manufacturers' national wholesale prices and rebates, both in dollar per ounce and their ratio. On average, the national wholesale price is \$1.032 per ounce, the rebate is \$0.858 per ounce, and the rebate-wholesale price ratio is 82.4\%, which is substantial. Nestl\'{e}'s rebate-wholesale price ratio is 86.4\%, higher than that of Abbott and MJ, whose ratios are 79.7\% and 82.5\%, respectively. Some of Nestl\'{e}'s rebates are even higher than its wholesale prices. For all three manufacturers, the rebate-wholesale price ratio is over 90\% for more than 25\% rebates submitted. In other words, the manufacturers only charge the WIC program 10\% of their national wholesale price for over 25\% of WIC contracts.

\section{Reduced-form evidence}
In this section, we present reduced-form evidence on the distortionary effects of the WIC program on the infant formula market. The analysis is motivated by the high rebates shown in Table \ref{table: sumstat of whole sale and rebate}, demonstrating that manufacturers compete aggressively for the WIC contracts. 




Intuitively, the aggressive competition for the WIC markets could be rationalized by the following features. First, the marginal cost for the manufacturers is very low, so it is still profitable to serve the WIC market with a high rebate. For this to be consistent with the significantly high retail price in the non-WIC market, non-WIC consumers need to be highly price insensitive.
Second, the marginal cost is not very low so the winner suffers a loss from winning, but there are additional benefits in both demand and supply to winning the WIC contract. Therefore, the additional benefit compensates for the loss, so the overall effect of winning is still desirable. Specifically, winning a WIC contract might boost the demand of non-WIC consumers and decrease the winner's marginal costs due to economies of scale, saving in advertisements, transportation, etc.  Since marginal costs are unobserved in the data, we provide some regression patterns to investigate how the equilibrium prices and sales change when the manufacturers are WIC winners.

\subsection{The effects of winning on volume sales}
\begin{figure}[h!]
\caption{Impacts of winning on sales}
\begin{multicols}{2}
    \includegraphics[width=\linewidth]{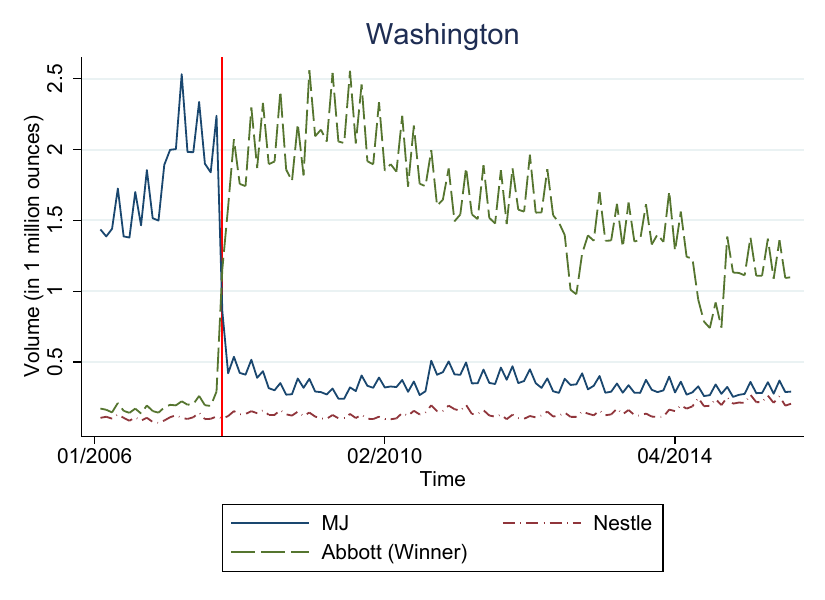}\par 
    \includegraphics[width=\linewidth]{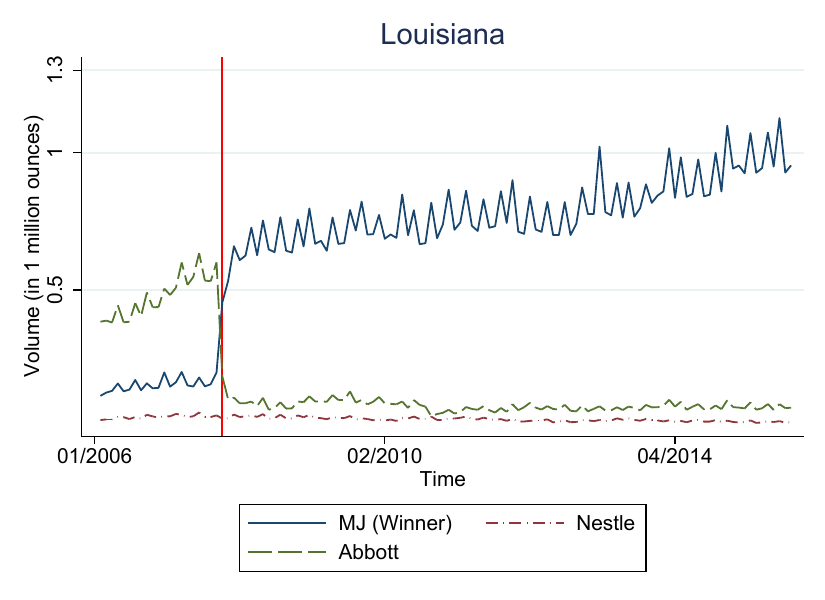}\par 
    \end{multicols}
\begin{multicols}{2}
    \includegraphics[width=\linewidth]{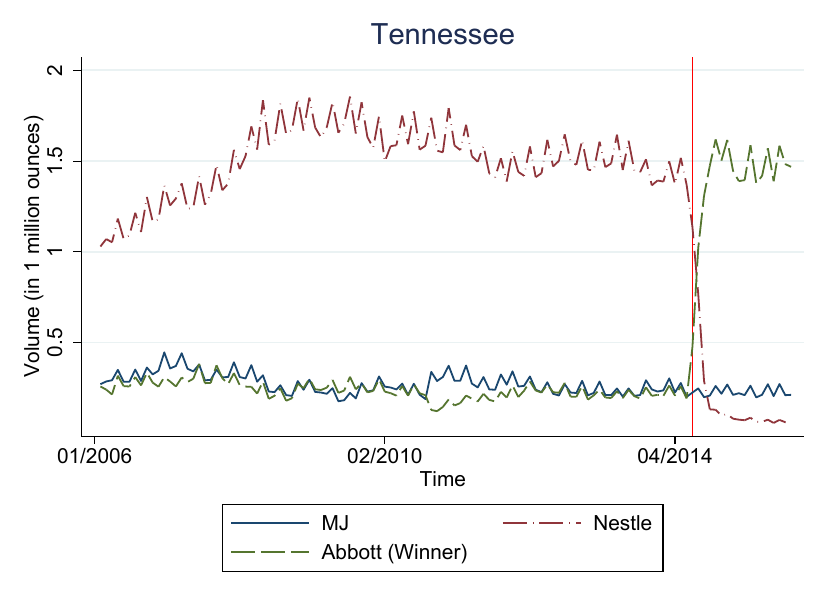}\par
    \includegraphics[width=\linewidth]{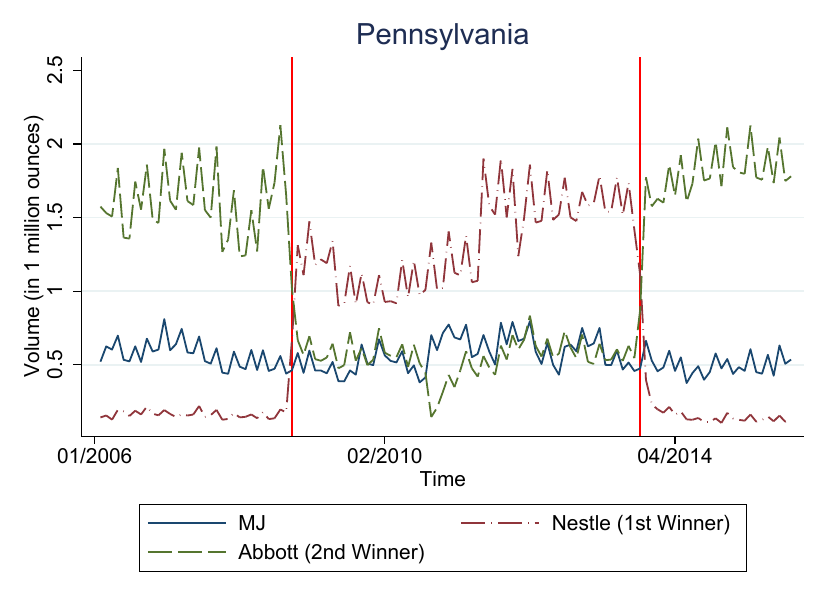}\par
\end{multicols}

\label{fig:impacts of winning on sales}\centering
\vspace{0.3cm}
\begin{minipage}{0.7\textwidth} 
{\footnotesize Note: The red vertical line indicates the starting time of a new WIC contract.}
\end{minipage}
\end{figure}

We first provide some visualization of the impact of winning a WIC contract on non-WIC sales. Figure \ref{fig:impacts of winning on sales} provides a visual illustration of the impact using Louisiana, Pennsylvania, Tennessee, and Washington as examples. The red vertical line indicates when the winner changes during the data period. For the three states except for Pennsylvania, the winner changed from MJ to Abbott, Abbott to MJ, and Nestl\'{e } to Abbott, respectively. For Pennsylvania, the winner switched from Abbott to Nestl\'{e} and then to Abbott. All four states display a similar pattern; we only describe it for Washington. MJ dominates the non-WIC infant formula market when it has a WIC contract. However, the volume sales of MJ plummeted when Abbott became the winner starting from October 2007, and the volume sales of Abbott in the non-WIC market skyrocketed immediately after the new contract began. The volume sales of Nestl\'{e} are relatively stable during the whole period.


To quantify the effect of winning WIC contracts on non-WIC sales,  we run a series of simple pooled regressions of the logarithm of volume sales on an indicator of winning and other covariates.
\begin{eqnarray}\label{eq:impact of winning on sales}
\log(V_{fjmt})&=&\beta_0+\alpha_m+\alpha_j+\sum_{f'}\alpha_{f'} I_{f'}+\beta_1 Win_{fmt}+\sum\nolimits_{f'}\gamma_{f'}  Win_{f'mt}\times I_{f'}\nonumber\\
&+&\beta_2 Auc_{fj}
+\beta_3 Win_{fmt} \times Auc_{fj}
+\sum\nolimits_{f'}\omega_{f'}  Auc_{fj}\times I_{f'}\nonumber\\
&+&\sum\nolimits_{f'}\zeta_{f'} Win_{fmt}\times Auc_{fj}\times I_{f'}
+\beta_4 \log(p_{fjmt})
+Z^{\prime}_{fmt}\xi+\epsilon_{fjmt},
\end{eqnarray}
where $f$ denotes manufacturer, $j$ denotes brand, $m$ denotes state, $t$ denotes month, $V_{fjmt}$ denotes the volume sales for manufacturer $f$'s brand $j$ in state $m$ at month $t$, $\alpha_m$ and $\alpha_j$ are state and brand fixed effects, respectively, $I_f$ is a manufacturer dummy, $Win_{fmt}$ is a dummy variable indicating whether manufacturer $f$ is the winner in market $m$ at year $t$, $Auc_{fj}$ is a dummy variable indicating whether brand $j$ is the auctioned  brand of manufacturer $f$, which does not vary across market $m$ and month $t$. $\log(p_{fjmt})$ is the logarithm retail price of brand $j$ for manufacturer $f$ in market $m$ and month $t$, $Z_{fmt}$ includes a variety of covariates, such as the number of non-WIC infants and other demographic variables that may affect the demand of infant formula. In equation (\ref{eq:impact of winning on sales}), $\beta_1$ captures the spillover effect, $\beta_3$ allows the spillover effect to be different for the WIC and non-WIC brands, and $\gamma_{f'},\delta_{f'},$ and $\rho_{f'}$ describes the variation of the spillover effects across manufacturers.

In the regression above, prices are potentially endogenous because they are from the equilibrium. We use raw milk price, the average distance between manufacturers' production centers to the market, and the electricity rate as instrumental variables for prices. These instruments are valid because they affect infant formula production and retail costs and, thus, are correlated with prices in the focal market; however, these factors are uncorrelated with market-specific demand shocks.

We present the regression results in Table \ref{tab:results_sales}. The main finding is that the non-WIC sales for the same manufacturers are much larger when winning than losing. The results in column (3) show that winning a WIC contract boosts the sales of non-auctioned and auctioned brands by 75\% and 126\%, respectively. The impacts of winning on sales are heterogeneous across manufacturers. For non-auctioned brands, the increase in sales for MJ is larger than that for Abbott and  Nestl\'{e}, but for auctioned brands, the rise in sales for MJ is smaller than that for Abbott and Nestl\'{e}. The larger benefit from winning for Nestl\'{e} and Abbott relative to MJ could be due to their relatively small market shares.



Several possible channels may lead to the above positive spillover effects. First, since the winning brand serves the WIC market exclusively, retailers adjust by allocating more shelf space and better product placement for this brand. Second, hospitals and physicians often recommend the winning brand considering that around half of the newborns are eligible for WIC. For example, the WIC agency in Wisconsin explicitly requires physicians to recommend the WIC brand.\footnote{Source: https://www.dhs.wisconsin.gov/publications/p4/p40023.pdf, retrieved on March 22, 2023.} Third, being the WIC brand may increase credibility for some non-WIC consumers, which might also serve as advertisements for WIC brands. Last but not least, WIC participants may affect non-WIC consumers through social networks or peer effects.

?\ref{tab:results_sales} presents the heterogeneity in the spillover effects across manufacturers. Overall, compared to Mead Johnson, Nestl\'{e} and Abbott benefits the most from winning the auctions. As a top seller in almost all markets,  Med Johnson will not experience a percentage increase in volume sales as high as as the other two manufacturers from winning the auctions.

\subsection{The effects of winning on retail prices}
Since WIC consumers are price inelastic, a retailer that serves non-WIC and WIC consumers could increase its profit by setting a higher price for the winning brand. Nevertheless, such an incentive may be dampened by the WIC regulation. Therefore, to investigate whether winning a WIC contract increases retail prices, we run a series of regressions of retail prices on a winning indicator and other covariates.
\begin{eqnarray}\label{eq:non-WIC sales_winning}
\log(p_{fjmt})&=&\beta_0+\alpha_m+\alpha_j+\sum_{f'}\alpha_{f'} I_{f'}+\beta_1 Win_{fmt}+\sum\nolimits_{f'}\gamma_{f'}  Win_{f'mt}\times I_{f'}\nonumber\\
&+&\beta_2 Auc_{fj}
+\beta_3 Win_{fmt} \times Auc_{fj}
+\sum\nolimits_{f'}\omega_{f'}  Auc_{fj}\times I_{f'}\nonumber\\
&+&\sum\nolimits_{f'}\zeta_{f'} Win_{fmt}\times Auc_{fj}\times I_{f'}
+\beta_4 \log(V_{fjmt})+Z^{\prime}_{fmt}\xi+\epsilon_{fjmt}.
\end{eqnarray}
where $Z_{fmt}$ is a vector of cost shifters, including the raw milk price, electricity rate, and distance between retailers and manufacturers. In the price regression above, the volume sale is endogenous because it is affected by price. We use demand shifters, such as the number of non-WIC infants, the logarithm of median income in a market, and the percentage of women labor participation in a market, as the instrument variables for the logarithm of volume sales. 

We present the regression results in Table \ref{tab:reduced_price}. There are several interesting observations from the results. First, we document a significant distortionary effect of the WIC contract on the price for non-WIC consumers. The results in column (3) demonstrate that the prices of winning manufacturers' auctioned brands are 8.1\% higher. However, winning a WIC contract does not significantly affect the price of the winner's non-auctioned brands. Note that the price of auction brands does not differ significantly from non-auction brands for losing manufacturers. Second, the effects of winning a WIC contract on prices are heterogeneous across manufacturers. For auctioned brands, the price increases for Abbott and Nestl\'{e} are 8.5\% and 7.9\% higher than MJ, as shown in column (6).

In summary, the regression results in this section document a significant distortionary effect of the WIC contract on the prices and quantities of the winner's brands. To further explore these sources, we propose a structural model of the market for infant formula in the next section.

\section{The Model}
This section presents a structural model concerning demand and supply for the US infant formula market. We first describe consumers' preferences and then propose a supply model where manufacturers compete sequentially for WIC contracts and retail pricing. In the retail pricing part, we further propose a two-step pricing game in that the winner determines the price for the WIC brand, and then all manufacturers do Bertrand pricing for the non-WIC brands. 

\subsection{Demand\label{section: demand}}

Suppose we observe $m=1, 2, \cdots, M$ markets, where a market is defined as a state-month combination.\footnote{In our analysis, we treat a multistate alliance as a state. Its demographics are taken average across the states in the alliance.} For ease of notation, Let $f$ and $f^*$ denote a generic manufacturer and the WIC auction winning manufacturer, respectively; $j$, $k$, and $k^*$ denote a generic, auctioned, and WIC (winning) brand, respectively; $\mathcal{N}_f$ denote the set of all the brands of manufacturer $f$, and $|\mathcal{N}_f|$ denote the number of brands of manufacturer $f$, $\sum_{f}|\mathcal{N}_f|=J$. 

Motivated by the reduced-form evidence that winning a WIC contract has spillover effects on the demand of non-WIC consumers, we model the indirect utility of a non-WIC consumer $i$ from product $j$ in market $m$ as follows.
\begin{eqnarray}\label{eq: utility function of consumers}
u_{ijm}= x_{jm} \beta_i- \alpha_i p_{jm}+ \delta_{0} \cdot\mathbbm{1}_{j\in \mathcal{N}_{f^{*}}}+\delta_{1}\cdot \mathbbm{1}_{j\in \mathcal{N}_{f^{*}}}\cdot \mathbbm{1}_{j=k^{*}}+\xi_{jm}+\varepsilon_{ijm},
\end{eqnarray}
where $\mathbbm{1}_{j\in \mathcal{N}_{f^{*}}}$ is an indicator that $j$ is a non-auctioned brand of the winning manufacturer $f^*$, $ \mathbbm{1}_{j=k^{*}}$ is an indicator that brand $k^*$ is the auctioned brand of $f^*$, $x_{jm}$ is a vector of observable product characteristics including manufacturer dummy,
$p_{jm}$ is the price of product $j$ in market $m$, $\xi_{jm}$ is the unobserved
product characteristics, and $\varepsilon_{ijm}$ is the idiosyncratic preference shock. The outside options for consumers are other formulas such as soy-based powder and other physical forms, store-brand infant formulas, e.g., Perrigo, and breastfeeding. 

In the specification above, we allow the spillover effects of winning a WIC contract on the auctioned brand ($\delta_{0}+\delta_{1}$) to be different from a non-auctioned brand ($\delta_{0}$).
If winning a WIC contract has no spillover effect, then $\delta_0=\delta_1=0$ in the utility specified in (\ref{eq: utility function of consumers}). Based on the reduced-form evidence, we expect that both spillover effects are positive, i.e., $\delta_0>0$ and $\delta_1>0$. The random coefficients $\beta_i$ and $\alpha_i$ may depend on consumers' demographic variables. For ease of exposition, we denote
 $p_m$, $x_m$, and $\xi_m$, respectively, the vector of prices, the vector of product characteristics, and the vector of unobserved product characteristics for all the brands in market $m$.

We further express the random coefficients as
 \begin{eqnarray}\left(\begin{matrix}
\alpha_i \\
\beta_i
\end{matrix}\right)= \left(\begin{matrix}
\alpha \\
\beta
\end{matrix}\right) +\Pi o_i+ \Sigma v_i,  v_i \sim P_v(v), o_i \sim  P_o(o),\end{eqnarray}
where $o_i$ is a vector of observed individual demographics, the matrix $\Pi$ captures how consumer characteristics $O_i$ affect taste, $v_i$ is a vector of unobserved individual characteristics, the matrix $\Sigma$ captures how consumer characteristics $v_i$ affect tastes, $P_v(v)$ is a standard multivariate normal distribution, and  $ P_o(o)$ is a known distribution from other data sources.

Each non-WIC consumer chooses a brand of formula or the outside option to maximize her utility. Note that the consumer's optimal choice depends on the winning manufacturer's identity because she prefers the brands of the winning manufacturers. We use $s^{f^*}_{jm}(p_m,x_m, \xi_m)$ to represent the market share of brand $j$ in market $m$ where the winning manufacturer is $f^*$. Following the existing literature in discrete choice demand estimation, we assume that the taste shocks $\varepsilon_{ijm}$ are i.i.d. draws from a type-one extreme value distribution so that the propensity of each household purchasing brand $j$ has a closed-form logit expression, resulting in the following aggregated market share representation.
 \begin{eqnarray}
 &&s^{ f^*}_{jm}(p_m, x_m, \xi_m) \nonumber\\
 &=&\int\frac{\exp(x_{jm} \beta_i- \alpha_i p_{jm}+ \delta_{0} \cdot\mathbbm{1}_{j\in \mathcal{N}_{f^{*}}}+\delta_{1}\cdot \mathbbm{1}_{j\in \mathcal{N}_{f^{*}}}\cdot \mathbbm{1}_{j=k^{*}}+\xi_{jm})}{\sum_{j'} \exp \left(x_{j'm} \beta_i- \alpha_i p_{j'm}+ \delta_{0} \cdot\mathbbm{1}_{j'\in \mathcal{N}_{f^{*}}}+\delta_{1}\cdot \mathbbm{1}_{j'\in \mathcal{N}_{f^{*}}}\cdot \mathbbm{1}_{j'=k^{*}}+\xi_{j'm}\right)}\mathrm{d} P_o(o)\mathrm{d} P_v(v),\nonumber \\
 \end{eqnarray}
 where the expectation is taken over the distributions of the random coefficients $\beta_i$ and $\alpha_i$. 
We then use $D^{f^*}_{jm}(p,x)= d_m  s^{f^*}_{jm}(p_m,x_m,\xi_m)$ to represent the demand of brand $j$ in market $m$ with $d_{m}$ being the size of market $m$ and the WIC winner being manufacturer $f^*$.

\subsection{Supply}

Based on the institutional background and the reduced-form evidence, we model the competition of the three manufacturers in market $m$ using a multi-stage game. Before the first stage, the national wholesale price for each manufacturer's auctioned brand (each manufacturer has only one) is realized and becomes common knowledge among all manufacturers.\footnote {Manufacturers could endogenously choose their auctioned brands. Nevertheless, auctioned brands are the most popular by sales, and they do not vary much across states or over time. Therefore, we assume auctioned brands are exogenously given.} 

In the first (auction) stage, there is an independent procurement auction in a market where manufacturers compete for the exclusive right to serve the WIC participants. Specifically, each manufacturer submits a rebate for its auctioned brand to maximize its overall expected payoff considering the auction outcome's uncertainty. Once the rebates are revealed, the exclusive right is awarded to the manufacturer with the lowest net price, calculated as the national wholesale price subtracting the submitted rebate. In the second (pricing) stage, each manufacturer sets retail prices for their products based on the auction winner's identity.\footnote{There are no auctions in some markets because a typical contract lasts for three years, and during the contract period, manufacturers only decide retail prices.}

We make the following two assumptions regarding the national wholesale prices to be consistent with the institutional background. First, the national wholesale price for the auctioned brand is non-binding in the post-auction competition. Even though the winner is determined jointly by the national wholesale price and the rebate, the government's expenditure, which is the difference between the retail price and rebate, does not depend on the national wholesale price after winning the auction. 
Second, the national wholesale price does not change in the post-auction period. Note that if the winning manufacturer does adjust its national wholesale price after winning the auction, the agency requires the winner to adjust the rebates by the same amount to keep the net price invariant. This unique feature by the agency discourages the winning manufacturer from raising the national wholesale price. 

We now present  the supply model backward and characterize its equilibrium. 
\subsubsection{The pricing stage\label{section: supply side}}
In each market, a manufacturer chooses brand-level retail prices to maximize its overall profit upon observing the auction winner's identity. For simplicity, we suppress the market index $m$. Note that the auction winners' and losers' pricing decisions differ, so we model them separately. 

Let $D^{f^*}_j(p)$ denote the demand of brand $j \in \mathcal{N}_f$ for  manufacturer $f$, where $p$ is the vector of prices of all the brands in the market,  and the superscript $f^*$ indicates the winning manufacturer's identity. The overall profit of manufacturer $f$ from the non-WIC market is
\begin{eqnarray}
\pi^{f^*}_{f}=\sum\limits_{j\in \mathcal{N}_f}(p_j-c_j)D^{f^{*}}_j(p)-C_f, 
\end{eqnarray}
where $c_j$ is the marginal cost of brand $j$, and $C_f$ is the manufacturer's fixed cost. We allow the non-WIC demand of brand $j$ for manufacturer $f$ at a given price vector $p$ to depend on the winner's identify $f^*$.\footnote{This is motivated by the reduced-form evidence that the impacts of winning on sales depend on the winner's identity.} If $f=f^*$, the profit function represents the winner's non-WIC profit.

The pricing competition in the infant formula market is complicated and unprecedented due to several unique features. First, there are two types of consumers, including regular non-WIC consumers and price-insensitive WIC consumers. If the winning manufacturer optimizes its overall profit, the optimal price would be set as high as possible because of the inelastic WIC consumers. Second, the WIC brand's price is regulated to prevent the winning manufacturer from discriminating against consumers and any retailer from charging much higher prices than  its peer retailers.\footnote{
The state agencies impose various restrictions on the pricing ceiling for the WIC products, based on the price of a retailer's peer group, including both non-WIC and WIC stores. Violating such a pricing ceiling might result in losing the WIC eligibility, so the pricing ceiling is binding.}  Third, winning may reduce marginal costs for the winner's products due to the economy of scale, saving in advertisement and transportation, etc.  

Facing these challenges, we model the pricing decision in two steps: the first is determining the WIC brand price, and the second is determining the non-WIC brand prices. In the first step (WIC pricing step), the winner determines the initial price for the WIC brand through a perceived profit maximization process where only non-WIC consumers are considered and assuming all other manufacturers choose prices to maximize their profits. This perceived process is essentially a Bertrand price competition.
\begin{eqnarray}\label{eq: perceived optimzation problem of the winner}
\tilde{\pi}^{f^*}_{f^*}&=&\sum\limits_{j\in \mathcal{N}_{f^*}/ {k^*} }\big(p_j-c_j\big)D^{f^*}_j(p) + \big(p_{k^*}-c_{k^*} \big)D^{f^*}_{k^*}(p) -C_{f^*},\nonumber\\
\tilde{\pi}^{f^*}_{f} &=&\sum\limits_{j\in \mathcal{N}_f}(p_j-c_j)D^{f^{*}}_j(p)-C_f, \hspace{0.3cm}f\neq f^*,
\end{eqnarray}
which generates a perceived equilibrium price for WIC brand, denoted as $\tilde{p}_{k^*}$. 

The winner then adjusts the WIC brand's price to respond to the demand from price-insensitive WIC consumers and comply with governmental regulations. Note that the WIC demand incentivizes the winner to raise the WIC brand price, but  governmental regulations restrict it from rising. Because of the implicit regulation, it is unclear how they affect the WIC brand's price. We combine all these factors and model the winner's price adjustment in the following reduced-form fashion:
\begin{eqnarray}
p_{k^*}=(1+\rho_{f})\tilde{p}_{k^*},
\end{eqnarray}
where $p_{k^*}$ is the final price of the WIC brand, $\rho_f$ is the price adjustment, which we refer to as the WIC pricing strategy and varies across manufacturers, i.e., $\rho_f$ is indicated by the manufacturer indicator $f$.

In the second step (non-WIC brand pricing step), taking the WIC brand price $p_{k^*}$ as given, all manufacturers compete to determine their retail prices for their products simultaneously in the traditional Bertrand fashion except the winner, who only solve the optimal price for its non-auctioned brands. The profit optimization for the winning and losing manufacturers can be summarized as follows.
\begin{eqnarray}
&&max_{j\in \mathcal{N}_{f^*}/ {k^*} } {\pi}^{f^*}_{f^*}=\sum\limits_{j\in \mathcal{N}_{f^*}/ {k^*} }\big(p_j-c_j\big)D^{f^*}_j(p_{-k^*},p_{k^*}) + \big(p_{k^*}-c_{k^*} \big)D^{f^*}_{k^*}(p_{-k^*},p_{k^*}) -C_{f^*},\nonumber\\
&&max_{j\in \mathcal{N}_{f}} {\pi}^{f^*}_{f}=\sum\limits_{j\in \mathcal{N}_f}(p_j-c_j)D^{f^{*}}_j(p_{-k^*},p_{k^*})-C_f, \hspace{0.3cm}f\neq f^*,
\label{eqn: wicb}
\end{eqnarray}
where $p_{-k^*}$ collects the vector of prices besides the price for the WIC brand. 
It is worth noting that there is a subtle difference in the winner's pricing optimization in this step from the standard Bertrand competition. Even though the winner does not adjust the WIC brand's price when the winner decides the price for her other brands, the winner still considers the negative effect of these prices on the WIC brand's non-WIC demand and, thus, the overall profit. 
Using the following first-order conditions, we can characterize the profit maximization separately for winning and losing manufacturers. A losing manufacturer $f$'s first-order-condition is
\begin{eqnarray}\label{eq: foc for loser in pricing game}
\frac{\pi^{f^*}_{f}}{\partial p_j}&=&\sum\limits_{j'\in \mathcal{N}_f}(p_{j'}-c_{j'})  \frac{\partial s^{f^{*}}_{j'}(p_{-k^*},p_{k^*})}{\partial p_j}+ s^{f^{*}}_j(p_{-k^*},p_{k^*})=0.
\end{eqnarray}
The first-order-condition of the winning manufacturer for its non-WIC brands is:
\begin{eqnarray}\label{eq: foc for winner in pricing game}
\frac{\partial \pi_{f^*}}{\partial p_{j}}&=&\sum\limits_{j'\in \mathcal{N}_{f^*}/ {k^*}}\big(p_{j'}-c_{j'}\big) \frac{\partial s^{f^*}_{j'}(p_{-k^*},p_{k^*})}{ \partial p_{j}}+ s^{f^*}_{j}(p_{-k^*},p_{k^*})\nonumber\\
&+& \big(p_{k^*}-c_{k^*} \big)\frac{\partial s^{f^*}_{k^*}(p_{-k^*},p_{k^*})}{ \partial p_{j}}=0, j \neq k^*.
\end{eqnarray}


The equilibrium conditions above are similar to that in the standard Bertrand pricing game framework, except that it does not consist of the optimization for the WIC brand. We can still represent the equilibrium pricing strategies in the following matrix expression:
\begin{eqnarray}
\label{eq: foc of manufacturers}
 \Delta S_{-k^*} \times \left( \mathbf{p} - \mathbf{c}\right)  + S_{-k^*} =0,
\end{eqnarray}
where $\mathbf{p}$ and $\mathbf{c}$ are $J\times 1$ vectors of prices and marginal costs, respectively with $J$ being the total number of brands in the market, including the WIC brand, and $\Delta S_{-k^*}$ is a $(J-1)\times J$ matrix defined as follows.
\begin{eqnarray}
\Delta S_{-k^*}(j,j') = \begin{cases}
  \frac{\partial s^{f^*}_j}{\partial p_{j'}},  & \text{$j$ and $j'$ are produced by the same manufacturer} \\
  0, & \text{otherwise}.
\end{cases}
\end{eqnarray}
In the definition above, $j'\neq k^*$ because the winning manufacturer does not set the price $p_{k^*}$ as in a Bertrand framework, so $\frac{\partial s^{f^*}_{j}}{\partial p_{k^*}}$ is excluded. $S_{-k^*}$ is a $(J-1)\times 1$ vector of market shares of all the brands except the WIC brand.

\subsubsection{The auction stage}
The procurement auction in which manufacturers compete to serve WIC infants exclusively is non-standard because of the following features. First, the payoff of a losing manufacturer may depend on the winner's identity due to the asymmetric competition between the three manufacturers, this is called identity-dependent externality. In contrast, the payoff of a losing bidder is usually zero regardless of who wins the auction in a standard auction. Moreover, the expected payoff of a manufacturer participating in the auction is much more complicated than in a standard auction: the payoff from winning depends not just on their costs but also on the opponents' costs because the cost affects the opponent's retail prices in the second stage. As a result, we cannot represent the expected payoff as a product of the benefit from winning and the probability of winning.
Lastly, the equilibrium strategy is asymmetric because manufacturers have different payoff functions due to consumers' heterogeneous preferences for brands.

Facing these challenges in modeling the manufacturers' bidding behavior in the WIC auction, even in theory, it is infeasible to show the existence and the uniqueness of equilibrium. Therefore, we propose to take an alternative route and rely on a flexible reduced-form analysis to approximate the determinant of the rebates from the data directly. The benefit of such an approach is that we do not need to take a stance on how the manufacturers view this complicated auction environment and save us from the risk of mis-specifications. However, the cost of such an approach limits our scope in conducting counterfactual analysis. We can only simulate the model outcomes limited to the situations where the data have some information about those situations. We will be more specific about the  limitations in the counterfactual analysis section.   

Intuitively, the rebates are affected by both the cost and demand shifters. Considering all the information available to a manufacturer, we use a predictive model to approximate the rebating rule that the manufacturers adopt for determining the rebate. We assume the bidding function is linear in a  manufacturer's wholesale prices and the distance to the production center, demand shifters, including income and the number of WIC infants and non-WIC infants, and cost shifters, including the raw milk prices and the electricity rate. The manufacturer also considers the competition from rivals, so the rivals' wholesale prices and transportation distances might affect the rebate, too.  

Note that there is a spill-over effect from winning the WIC auction to the non-WIC market, captured by the boosted consumer preferences for the winning manufacturer's products, i.e., $\delta_{0}$ and $\delta_{1}$. The manufacturers' rebate could vary with such a spill-over effect. However, such a spill-over effect results in a heterogeneous spill-over impact in sales, depending on the market characteristics, including the cost and demand shifters. As a result, it is infeasible to quantify the direct consequence of the spill-over effect on the rebate. This infeasibility limits our capability to simulate the retail prices in the scenario varying the spilled-over effect. That is, the specification of the reduced-form regression of the rebate implicitly reveals the impact of the spill-over effect on the rebate. Our counterfactual analysis, therefore, can only be conducted with the assumption that the spill-over effect is the same as in the data.

\section{Identification and Estimation}

In this section, we first describe the identification and estimation of both the demand and supply sides and then present how we predict the rebate in the auction stage. We focus on identification and estimation on the supply side because of the challenges discussed in the last section,  while the demand side is standard following the seminal work BLP. 

\subsection{Identification}

It is worth emphasizing that we allow the manufacturers' marginal costs to depend on whether they win the WIC auction. Specifically, we model the marginal costs $c_{jm}$ as a linear function of product characteristics and cost savings:
\begin{eqnarray}
{c}_{jm}= Z_{jm} \gamma + \Delta c_{-k^*} \cdot\mathbbm{1}_{j\in \mathcal{N}_{f^{*}}}+\Delta c_{k^*} \cdot \mathbbm{1}_{j\in \mathcal{N}_{f^{*}}}\cdot \mathbbm{1}_{j=k^{*}} +\omega_{jm},\end{eqnarray}
where $Z_{jm}$ collects manufacturer and auction brand level fixed effects, product characteristics such as 
spitup and prebiotics, and cost shifters such as transportation distance, the raw milk price, and the electricity price. $\omega_{jm}$ is idiosyncratic shock to marginal cost.   

The model parameters from the supply side include marginal costs for all the brands in the market $c_j, j=1, 2, \cdots, J$, cost reductions of the winning manufacturer for the WIC brand $\Delta c_{k^*}$ and the non-WIC brands $\Delta c_{-k^*}$, and the price adjustment coefficient $\rho_f$. 
Identifying parameters on the supply side is challenging and new because our model in Section \ref{section: supply side} differs from a conventional Bertrand pricing competition, where one can recover the marginal cost from the manufacturers' first-order conditions. 

We take multiple steps to recover the parameters. First, considering that the pricing strategies of losing manufacturers are the same as in a Bertrand pricing model, we use the observed prices 
and demand functions to identify the $J-|\mathcal{N}_{f^*}|$ marginal costs for the losing manufacturers from the following first-order-conditions. 
\begin{eqnarray}
 \Delta S_{-k^*} \times \left( \mathbf{p} - \mathbf{c} \right)  + S_{-k^*} =0,
\end{eqnarray}
where only the $J-|\mathcal{N}_{f^*}|$ conditions for the losing manufacturers are used. 

Second, we establish that the marginal costs of non-WIC brands can be expressed as a linear function of the WIC brand's marginal cost. Note that it is infeasible to recover the winning manufacturers' marginal costs from its first-order-conditions because there are $|\mathcal{N}_{f^*}|$ marginal costs but $|\mathcal{N}_{f^*}|-1\equiv\ell$ equations, because the winner only optimizes for its $\ell$ non-WIC brands but not the WIC one. For ease of exposition, assume that brand 1,...,$\ell$ are  non-WIC brands.
For the $\ell$ first-order conditions of the winner described in Equation \ref{eq: foc for winner in pricing game}, we can represent the marginal cost for each of the non-WIC brands as a linear function of the marginal cost of the WIC brand. 
\begin{eqnarray}
    \begin{pmatrix}
    c_1\\
    c_2\\
    \vdots\\
    c_{\ell}
    \end{pmatrix}
    =    \begin{pmatrix}
    p_1\\
    p_2\\
    \vdots\\
    p_{\ell}
    \end{pmatrix}
    +\begin{pmatrix}
    \frac{\partial s^{f^*}_{1}}{ \partial p_{1}}& \frac{\partial s^{f^*}_{2}}{ \partial p_{1}} & \cdots \frac{\partial s^{f^*}_{\ell}}{\partial p_{1}}\\
    \frac{\partial s^{f^*}_{1}}{ \partial p_{2}}& \frac{\partial s^{f^*}_{2}}{ \partial p_{2}} & \cdots \frac{\partial s^{f^*}_{\ell}}{\partial p_{2}}\\
    \vdots\\
       \frac{\partial s^{f^*}_{1}}{ \partial p_{\ell}}& \frac{\partial s^{f^*}_{2}}{ \partial p_{\ell}} & \cdots \frac{\partial s^{f^*}_{\ell}}{\partial p_{\ell}}\\
    \end{pmatrix} ^{-1} 
    \left\{
        \begin{pmatrix}
            s^{f^*}_1\\
            s^{f^*}_2\\
            \vdots\\
            s^{f^*}_{\ell}
        \end{pmatrix}
        + \big(p_{k^*} -c_{k^*} \big)\cdot
        \begin{pmatrix}
            \frac{\partial s^{f^*}_{k^*}}{ \partial p_{1}}\\
            \frac{\partial s^{f^*}_{k^*}}{ \partial p_{2}}\\
            \vdots\\
            \frac{\partial s^{f^*}_{k^*}}{ \partial p_{\ell}}\\
        \end{pmatrix}
    \right\}.
\end{eqnarray}
Rewrite the matrix equation above,  
\begin{eqnarray}\label{equation: marginal costs of winner's non-auction brands}
c_{-k^*} = a_{-k^*} + b_{-k^*} c_{k^*},
\end{eqnarray}
where $a_{-k^*}$ and $b_{-k^*}$ are $\ell \times 1$ vectors and can be calculated directly from the data. Therefore, the marginal costs for the non-WIC brands can be non-parametrically identified and estimated once the marginal cost of the WIC brand is known. 

Third, we identify the marginal cost of the WIC brand using the winning manufacturer's pricing decision. The identification is achieved for any fixed parameter of price adjustment $\rho_{f^*}$. That is, we show that the marginal costs of all the brands produced by the winning manufacturer can be identified as a function of the parameter $\rho_{f^*}$. We will discuss how to calibrate $\rho_{f^*}$. For a given $\rho_{f^*}$, the (unobserved) perceived optimal price for the WIC brand is 
$$\tilde p_{k^*}=\frac{1}{(1+\rho_{f^*})}{p}_{k^*},$$
where ${p}_{k^*}$ is observed in the data, indicating that the perceived optimal price for WIC brand considered only WIC demand is known once $\rho_{f^*}$ is given. Note that the perceived optimal price $\tilde p_{k^*}$ is determined through the following Bentrand pricing optimization:
\begin{eqnarray}
 \Delta \tilde {S}\times \left( \mathbf{\tilde{p}} -  \mathbf{c} \right) + \tilde {S} =0,
\end{eqnarray}
where $\Delta \tilde {S}$ and $\tilde {S}$ are defined analogously to $S_{-k^*}$ and $\Delta S_{-k^*}$, respectively, but the row for the WIC brand was added back, such that $\Delta \tilde {S}$ is a $J\times J$ matrix, and  $\tilde {S}$ is a $J \times 1$ vector. Vector $\mathbf{c}$ consists of all the marginal costs for all products in the market, with losing manufacturers' marginal costs being known, and winning manufacturer's marginal costs of non-WIC brands are a linear function of the marginal cost of the WIC brand as specified in equation (\ref{equation: marginal costs of winner's non-auction brands}). 

If the perceived optimal prices are observed in the data, the first-order equations above construct a system of equations with the WIC brand's marginal costs being the only unknown. We can solve the equations to obtain the marginal cost of the WIC brand $c_{k^*}$ as a function of the parameter $\rho_{k^*}$. However, one only observes the perceived optimal price for the WIC brand. Intuitively, A vector of the marginal cost will generate a vector of equilibrium price. Therefore, one can use the observed $\tilde p_k$ to pin down the unknown $c_{k^*}$. Consequently, all marginal costs for non-WIC brands of the winning manufacturer can be recovered as functions of the parameter $\rho_{k^*}$, too.

Once we recover all the marginal costs nonparametrically for any given $\rho_{k^*}$, we can directly identify the cost savings $\Delta c_{-k^*}$ and $\Delta c_{k^*}$, which are components of marginal costs for non-WIC and WIC brands of the winning manufacturer, respectively. Intuitively, 
after controlling for other factors that affect marginal costs, the difference in marginal costs between a market where the manufacturer loses and a market where the manufacturer wins allows us to obtain cost savings. A potential issue of such an approach is that the cost shocks might have different distributions in the case of winning and losing. The manufacturer might be subject to more negative shocks than losing in those winning markets, leading to lower marginal costs and higher winning probabilities. Nevertheless, this is less likely in our case because the manufacturers bid once every three years (it could be longer oftentimes or shorter occasionally between two auctions in a market). At the same time, the post-auction pricing competition takes place on a monthly level. It is unlikely for the manufacturers to predict the cost shocks three years ahead. 

Note that the identification of the marginal costs and cost savings is up to the WIC pricing of the winning manufacturer, $\rho_{k^*}$. We now provide additional information to identify the WIC pricing strategy parameter. First, Such a pricing strategy is proposed to reduced-formally characterize the two special features in our model: perfectly price inelastic WIC demand and governmental regulation; both work in opposite directions and sort of cancel each other out. Failing explicitly to model their roles in pricing decisions, it is infeasible to point identify the parameter without additional information. As discussed in Section \ref{section: market and data}, data reveal that the three manufacturers bid aggressively to win the exclusive right to serve the WIC consumers. As a rational agent, the manufacturer would not want to bid or even win if the expected profit from losing is higher than from winning. This feature provides additional restrictions on the WIC pricing strategy. We use this intuition to identify a lower bound of the parameter at which a manufacturer is different between winning and losing, i.e., the expected profit of winning is the same as losing. Specifically, the WIC pricing strategy should satisfy the restriction that the winner's profit is not smaller than that if the manufacturer loses, which provides a set identification of the WIC pricing strategy.

\subsection{Estimation}
In this section, we present the estimating strategies for the model parameters. First, the demand parameters are estimated using the generalized method of moments (GMM) as in BLP. We then estimate the supply model following the identification procedure closely. 

\subsubsection{Demand estimation}
The estimation is to exploit population moment conditions constructed by the product of the structural error $\xi$ and the instrumental variables. Specifically, Let $\theta_D$ collect all preference parameters, i.e., $\theta_D \equiv\{\alpha, \beta, \delta_0, \delta_1, \Sigma, \Omega\}$, and let $\delta_{jm}$ represent the mean utility of brand $j$ in market $m$, that is,
\begin{eqnarray*}
\delta_{jm}=y_{jm} \beta- \alpha_i p_{jm}+ \delta_{0} \cdot\mathbbm{1}_{\{j\in \mathcal{N}_{f^{*}}\}}+\delta_{1}\cdot \mathbbm{1}_{\{j\in \mathcal{N}_{f^{*}}\}}\times \mathbbm{1}_{\{j=k^{*}\}}+\xi_{jm}.
\end{eqnarray*}
In the equation above, the price $p_{jm}$ is likely to be affected by the unobserved
characteristic $\xi_{jm}$. We use instrumental variables to control for the endogeneity following the existing literature. 
Instruments used include three sets: (1) cost shifters, including raw milk price, electricity price, and the average transportation distance from a manufacturer's production plant to a market, (2) BLP instruments, including characteristics of rivals' products in the same market, and (3) Hausman price IVs,  i.e., prices of products in other markets. 

We construct moment conditions using those instrumental variables. 
Let $V_{k,jm}$ be the $k$-th instrumental variable for brand $j$ in market $m$, we have $E[\xi_{jm} V_{k,jm}]=0$. The sample analog of the moment condition from the instrument $V_{k,jm}$ can be represented as
\begin{eqnarray*}
g_{k}(\theta_D)= \frac{1}{nobs}\sum_{m,j} \left[y_{jm} \beta- \alpha p_{jm}+ \delta_{0} \cdot\mathbbm{1}_{\{j\in \mathcal{N}_{f^{*}}\}}+\delta_{1}\cdot \mathbbm{1}_{\{j\in \mathcal{N}_{f^{*}}\}}\times \mathbbm{1}_{\{j=k^{*}\}} \right]V_{k,jm}, \end{eqnarray*}
where $nobs$ is the overall number of brands across markets and time. 
Stacking all the moment conditions induced by all the instrumental variables and denoting them as vector $g(\theta)$, the GMM estimator of the parameters $\theta_D$ is
\begin{eqnarray}
\hat \theta_D \equiv \arg \min\nolimits_{\theta} g(\theta)' A^{-1} g(\theta),
\end{eqnarray}
where $A$ is an estimate of the efficient weighting matrix based on parameter estimates obtained from a first-stage estimation with a 2SLS weighting matrix. 

\subsubsection{Supply estimation} 
The estimation on the supply side takes several steps. First, we nonparametrically estimate the marginal costs of all the brands for losing manufacturers in any single market using the first-order conditions of their profit optimization problems. 

In the second step, for a given pricing adjustment, we nonparametrically estimate the marginal cost of winning manufacturers by using her optimization condition, which involves the optimal pricing for non-WIC brands and the determination of the WIC brand's price. Specifically, given the marginal cost of the WIC brand, we recover the marginal cost of the non-WIC brand nonparametrically. We then can estimate the WIC brand's marginal cost by finding such a market-level marginal cost so that the perceived optimal price for the WIC brand is the same as that observed in the data with the pricing adjustment. Once the marginal costs are non-parametrically estimated, we exploit variation of estimated marginal costs across states to estimate the cost parameters, including the cost-saving parameter $\Delta c_{-k^*}$ and $\Delta c_{k^*}$. 

Lastly, we pin down the manufacturer-specific WIC price adjustment $\rho_f$. Note that the WIC pricing parameters are only partially identified, and estimating the identified set via moment inequality is challenging. Fortunately, the profit from winning is monotone with the WIC pricing strategy, given other factors. Therefore, we can estimate the lower bound of manufacturer-specific $\rho$ such that a manufacturer's profit from winning is the same as that from losing. Instead of searching for all possible values, to ease the computation burden, we allow the WIC pricing strategy for each manufacturer to be chosen from any point in the predetermined set $\{0.01, 0.02,...,0.2\}$ and choose the combination of the set such that the average profit from winning is not lower than that from losing for all three manufacturers. 

In particular, for each given set of $\rho$ for all three manufacturers, we first directly compute the WIC winner's profit in each market using the recovered marginal costs, prices in the data, market shares, and the number of WIC and non-WIC infants. Second, we calculate the expected profit of the winner from losing by assuming an equal probability that each opponent wins, where we solve for the new equilibrium price and market shares for the hypothetical new winner. Note that only the winner benefits from a cost saving and the demand spill-over effect, so we adjust the marginal cost for the losing and winning manufacturers accordingly before we solve for the new equilibrium. We compute the profit under such a hypothetical situation. We have to conduct these exercises for every market.

\subsection{Reduced-form Bidding strategies}
To estimate the reduced-form bidding (rebate) function, we assume rebates to be a linear function of three categories of variables. The first category is cost shifters, including manufacturer fixed effects, the distance between a manufacturer's production plant to a market, which captures the transportation cost, raw milk price, and electricity price. The second category is demand shifters such as the length of the contract, the number of WIC and non-WIC infants, and the median state-level income in a market. The third category of variables includes wholesale price and distance of rivals that account for a manufacturer's strategic response to its rivals in the bidding process. Since each manufacturer has two rivals, we take an average of the variables for two rivals to represent the overall competition in the market. 

In the data, we only observe two participating manufacturers in some auctions. MJ and Abbott participate in almost all WIC auctions while Nestl\'{e} does so less frequently. Nevertheless, since all three manufacturers have competed in the market for a long time, it is reasonable to assume that all the manufacturers believe that the other two manufacturers will participate in the auction and submit a rebate.

\section{Estimation Results}

\subsection{Demand}
We present the estimation results for the main demand specification in Table \ref{table: demand}. The estimated means of the distribution of marginal utilities are summarized in the first column. All coefficients except the constant are statistically significant and of the expected sign. The positively significant estimate of the winning dummy variable implies that consumers prefer products of the winning manufacturer, i.e., there are significant spillover effects of winning a WIC contract on \textit{all} the products of the winner. Moreover, there are additional significant 
spillover effects on the WIC brand of the winner on top of other brands of the same manufacturer. 
The estimated coefficients on the manufacturer-level fixed effects demonstrate that consumers prefer products of MJ the most, then of Abbott and Nestl\'{e} the least. This is consistent with the observed market shares in the markets. 

The next two columns present estimates of preference heterogeneity of the means in the first column. The standard deviation estimates for price and constant are insignificant, implying that marginal utility heterogeneity is not substantial for infant formula consumers. However, the estimates of 
the interaction between price and demeaned income is significantly positive, suggesting that high-income consumers are less sensitive to price. 

We estimate the implied own- and cross-price elasticities of demand among the three manufacturers and between WIC and non-WIC brands for all the markets and present the median elasticities in Table \ref{table: elasticities}. Each entry $i$,$j$, where $i$ indexes row and $j$ column, gives the elasticity of brand $i$ with respect to a change in the price of brand $j$. The own-price elasticities are negative for all products. The median elasticity across markets and brands is roughly -2.5, smaller than other food products studied in the literature, such as breakfast cereal estimated in \cite{nevo}. This relatively low price sensitivity is consistent with the fact that parents are usually reluctant to switch 
infant formula products even if the price rises. In general, consumers are less elastic for auctioned brands than for non-auctioned brands for all three manufacturers. Consumers' demand is the least elastic for Abbott's auctioned brand and most elastic for MJ's non-auctioned brand.  

Moreover, our estimated substitution patterns exhibit significant variation across auction and non-auction brands. Abbott's auction brand appears to be the most popular substitute for all other products, MJ's auction brand is the next, and Nestl\'{e}'s auction brand is the least popular. The auction brand of MJ is a closer substitute of Abbott's than Nestl\'{e}'s, suggesting that consumers view MJ and Abbott's products similarly but view Nestl\'{e}'s products differently from the other two manufacturers. This finding is consistent with the fact that Nestl\'{e}'s price is relatively lower than the other two manufacturers.

\subsection{Supply\label{sec: estimation of supply side}}
Using the demand estimates, we estimate the marginal costs of all the products nonparametrically from the first-order conditions of losing and winning manufacturers' optimization problem for a given WIC pricing strategy and estimate the cost saving and cost functions using those marginal costs. We present the estimation results in Table \ref{pricing1}.\footnote{Supply parameters are robust to variation of the WIC price adjustments from 0.01 to 0.2.} 

The results display a few interesting patterns. First, there is considerable heterogeneity in price adjustment across the three manufacturers. Nestl\'{e} increases the WIC brand's price by 11\% by the perceived optimal price without considering WIC demand. In contrast, the increases for Abbott and MJ are 7\% and 1\%, respectively. This can be due to the fact that Nestl\'{e} charges a relatively lower price for its products without a WIC contract. All cost savings are statistically significant at 1\%.

Second, there is also massive heterogeneity in the cost savings from winning. MJ's cost saving for its WIC brand is only slightly larger than its non-WIC brands. Compared with MJ, Nestl\'{e} has a lower cost saving for its non-WIC brands, but a higher cost saving on the WIC brand; Abbott has similar cost saving to MJ.  

\begin{figure}[tbp]
\caption{Histogram of estimated marginal costs and markups}
\centering
\label{fig:marginalcost} 
\includegraphics[height=3.6in]{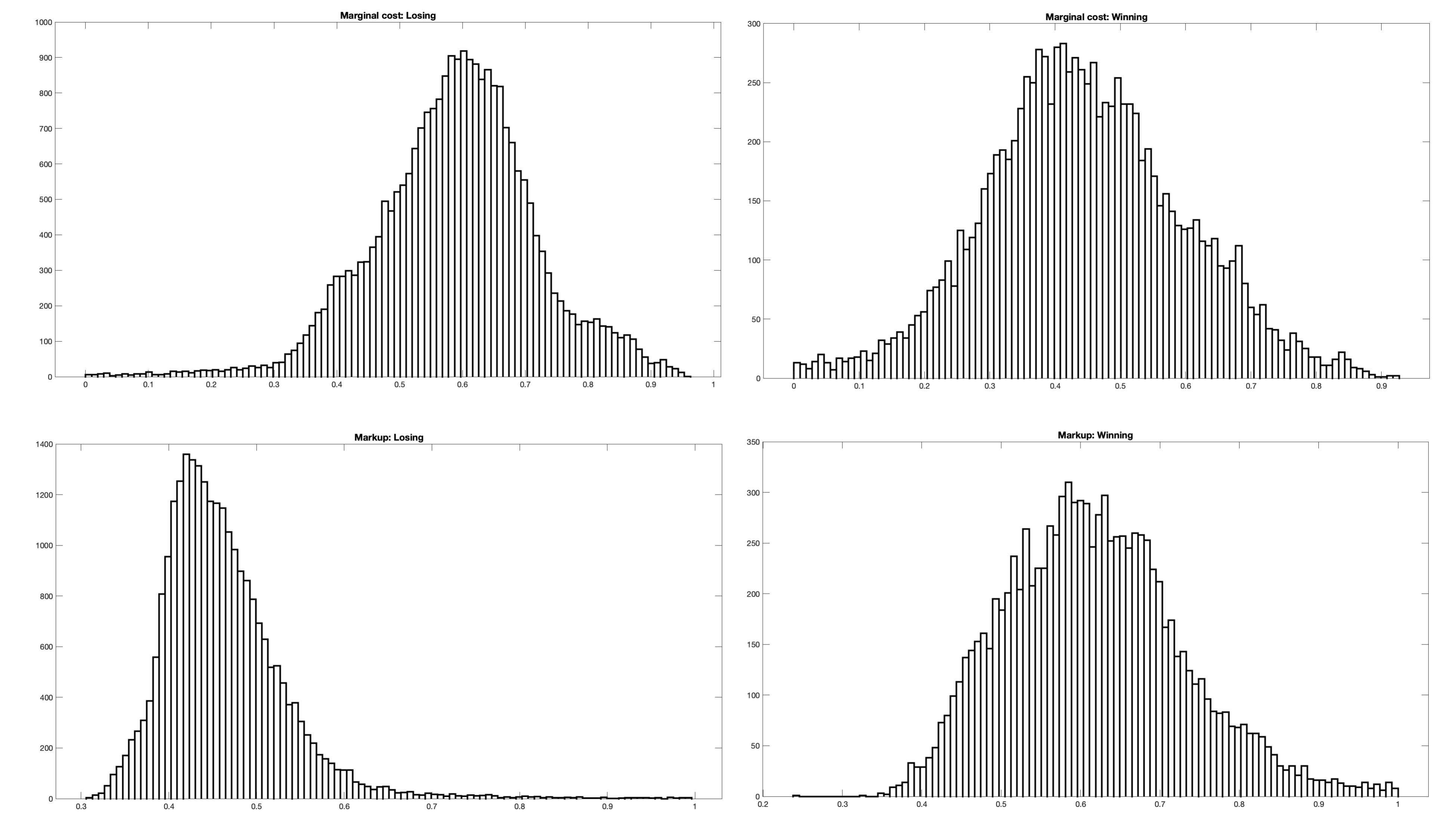}
\end{figure}

We present the histograms of the estimated marginal costs and markups in Figure \ref{fig:marginalcost} and their summary statistics in Table \ref{est: markup} separately when winning and losing WIC auctions. When manufacturers do not have a WIC contract, the mean and median marginal costs of all products across all markets are estimated to be \$0.587/ounce and \$0.594/ounce, respectively. The markup has a median of 45\% and a mean of 46\%, with a small standard deviation of 8.5\% across all products and all markets. The relatively high markups of informant formula products are due to the considerable market power of the three manufacturers who dominate the market. Because of cost savings due to winning a WIC contract, the marginal cost is lower and the markup is higher upon winning for all three manufacturers. 

The histogram in Figure \ref{fig:marginalcost} and the estimated results in Table \ref{est: markup} illustrate significant heterogeneity of marginal costs and markups across three manufacturers. The heterogeneity of marginal costs across manufacturers is relatively substantial: MJ's marginal cost is the highest (\$0.636/ounce) on average, and Nestl\'{e} is the lowest (\$0.569/ounce) with a 10.5\% difference. However, the markups of the three manufacturers are more uniform; the highest markup (Abbott) is only 4.4\% higher than the lowest markup (Nestl\'{e}). The increase in markup due to winning a WIC contract also differs. Abbott's mean markup rises from 47.8\% to 67.0\%, a 40.2\% increase, while the change is 30.4\% for MJ and 31.5\% for Nestl\'{e}. 

\begin{figure}[tbp]
\caption{Histogram of manufacturers' profit}
\centering
\label{fig:profit} 
\includegraphics[height=3.6in]{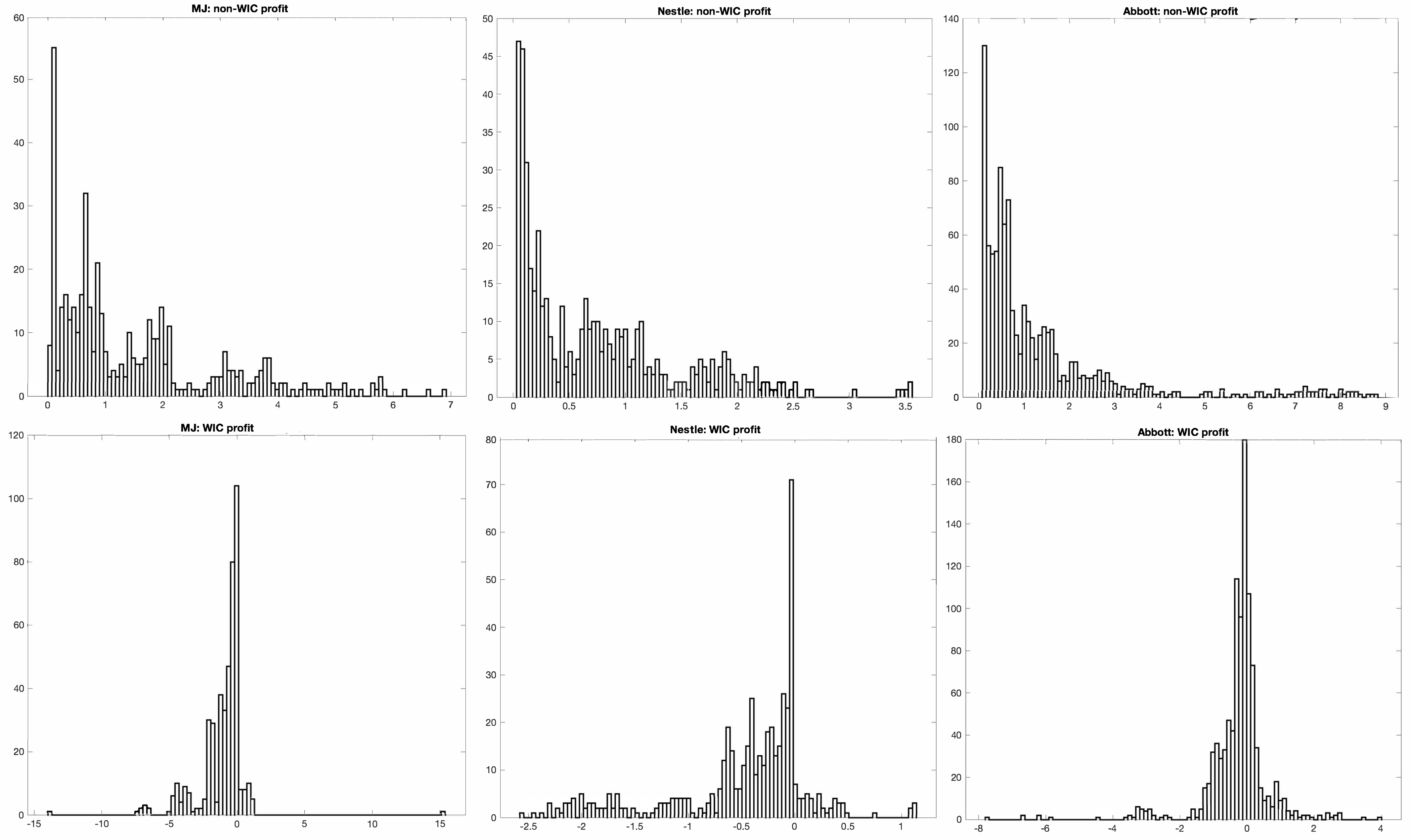}
\end{figure}

Figure \ref{fig:profit} illustrates the distributions of manufacturers' profits in both the WIC and non-WIC markets. The average profits in a WIC market are -1.044, -0.482, and -0.265 million dollars for MJ, Nestl\'{e}, and Abbott, respectively. All three manufacturers have negative profits from more than half of the WIC markets. 

Our findings have an interesting implication for the structure of the infant formula market: if a manufacturer decides to participate in a WIC auction in a market, then a sufficient large non-WIC market share is necessary for the manufacturer to survive. Suppose the manufacturer wins in a market, then its profit in the WIC market is negative on average due to the high rebate. The manufacturer has to get a sufficiently large profit from the non-WIC market. On the other hand, if the manufacturer loses the auction, it will also lose a substantial portion of its demand to the winner due to the spillover effects. This indicates that the presence of the exclusive right by auction might affect the manufacturers asymmetrically in the market.

\subsection{Reduced-form Bidding functions\label{sec: estimation of bidding function}}

We estimate the approximated bidding functions under several specifications and present the estimation results in Table \ref{table: rebate}. A few interesting patterns arise. First, a manufacturer's rebate responds positively and statistically significantly to its wholesale price: every percent increase in the wholesale price leads to a 2.1\% increase in rebate. This suggests that the higher the wholesale price, the higher the rebate to maintain the profit margin. Moreover, a manufacturer submits a 1.1\% lower rebate if its rival's wholesale price is 1\%  higher. This negative association is statistically significant for the three more complete specifications (columns (3)-(6)). This effect is the outcome of the manufacturer weighing its winning probability and payoff. The rival would submit a higher rebate if its wholesale price is higher. Therefore, the probability of winning is smaller for a manufacturer. The manufacturer could bid higher to increase the winning probability. However, given a winning probability, the manufacturer would bid lower to make more profit. 

The estimates also show that a manufacturer bids a lower rebate in a market with a larger number of WIC infants. This can be rationalized by the fact that the winning manufacturer loses money in the WIC market. Thus the returns from winning are lower. By contrast, the rebate increases in the non-WIC market size. This is mainly due to the substantial spillover effects of winning the WIC contract. 


Finally, we also observe that everything else equal, Nestl\'{e}'s rebate is 9\% higher than MJ, and the difference is statistically significant. Meanwhile, the difference in rebates between Abbott and MJ is not statistically significant. This demonstrates that Nestl\'{e} bids more aggressively than its opponents to maintain or increase its market power.

\section{Counterfactual Analyses}     
In this section, we aim to evaluate the effectiveness of government procurement fully. First, what are the impacts of the WIC procurement auction on the government, consumers, and manufacturers? Second, how WIC program size affects the market outcomes, i.e., government expenditure, consumer surplus, and manufacturer profits? Lastly, what would the market outcome be if the government adopts a pre-predetermined rebate approach employed by MDRP? For these purposes, we simulate the outcomes under three common procurement methods: (1) competitive bidding as implemented in the current infant formula market, (2) the government directly reimburses the program participants' purchases, as in the case of other food products in WIC, and (3) the government sets a predetermined rebate and those manufacturers who choose to participate have to provide the rebate. 

\subsection{Basic setup}
The first issue we need to address in our simulation is that some key market features, including consumers' spill-over effects for WIC and non-WIC consumers, cost savings from winning the auctions, and the WIC pricing strategy, are estimated using data with the WIC auction. Those features may differ in the alternative mechanisms, so we must make assumptions about those features in our simulations. 

When the government reimburses the program participants directly without competitive bidding, we assume that the government still has some requirements on the products provided by WIC and only the current auction brands, which are the most popular products, are available for WIC consumers. Since there is no sole WIC provider anymore, we assume that the WIC manufacturers share the spill-over effect equally. For instance, if all three manufacturers provide WIC products, the spill-over effect becomes 1/3 of the original estimated effect in the competitive bidding mechanism. Under this assumption, non-WIC consumers prefer WIC products to non-WIC ones, but the magnitude becomes 1/3 of that under the competitive bidding mechanism. We make similar assumptions for cost savings. Lastly, because WIC manufacturers are still facing two market segments, one being price inelastic, we assume that manufacturers adopt the same pricing strategy as that with auctions. That is, all WIC manufacturers solve for the optimal prices for their WIC products considering only the non-WIC profits. The manufacturers then determine their WIC products' prices using their WIC pricing strategies. Given the WIC product prices, they then optimize their prices for non-WIC products. Analogously, we also make the same assumptions above in the pre-determined rebates mechanism.

Another challenge to our simulation study is the lack of 
information about WIC consumers' preferences. To address this issue, we assume that WIC consumers have the same preference as non-WIC consumers over infant formulas. Under such an assumption, WIC consumers would choose the brand that maximizes their indirect utility, but the coefficient of the price in their indirect utility is set to zero.

To simulate the market outcome, we use the information of all the markets in 2015 from the data (there are 502 markets in total). A market's exogenous features include demand and supply shifters, for instance, the number of WIC and non-WIC infants, demographics in the state, the manufacturers' cost shifters, wholesale prices, etc.

\subsection{Procurement mechanisms}
In this subsection, we provide details on the three procurement methods and how we solve for equilibrium for each method.

\subsubsection{Competitive bidding}
When competitive bidding is adopted as the procurement mechanism, we  simulate rebates to determine the winner in any market with given market features. 
The reduced-form bidding function allows us to predict the rebate for a given set of market characteristics. To fully capture the rebate, we randomly draw residuals of the bidding function in Section \ref{sec: estimation of bidding function} and add them back to the predicted rebate, with which we simulate the market outcomes. For each market, we randomly draw residuals 50 times. Once we obtain the simulated rebates for all manufacturers, the winner is determined as the manufacturer with the lowest net price.

Using the simulated rebate and winner identity in a given market, we solve the optimal retail prices for all the brands in two steps. First, the winner determines the price for the WIC brand using the first-order conditions (\ref{eqn: wicb}). Once the WIC brand's price is determined, winning and losing manufacturers solve for optimal prices for other brands using the first-order conditions (\ref{eq: foc of manufacturers}). The pricing equilibrium is solved by iterating on the manufacturers' best response until convergence.  After obtaining the equilibrium retail prices, we can compute the WIC and non-WIC consumers' surplus, manufacturers' profits, and government expenditure.

\subsubsection{Procurement without competitive bidding}
The government can procure without taking explicit measures to contain the cost. This is common in social welfare programs, e.g., the Supplemental Nutrition Assistance Program (SNAP) program sets eligibility criteria for participating vendors. In this section, we examine outcomes under such a procurement method. 
In such a mechanism, WIC consumers may choose any WIC brands they prefer, and the government will pay for them. Just as in competitive bidding, manufacturers still set retail prices by considering two segments of consumers, and the government regulates the prices as in the competitive bidding mechanism. Under the assumptions above, we solve for the optimal retail prices, then compute the WIC and non-WIC consumers' surplus, manufacturer profit, and government expenditure. 

\subsubsection{Predetermined rebate programs}
Another prominent procurement method is pre-determined rebates adopted by MDRP. The federal government and the states share Medicaid expenditures and rebate savings. States reimburse pharmacists directly for drugs purchased by Medicaid beneficiaries and then report their expenditures to the federal government for partial reimbursement.
Under the basic rebate formula, pharmaceutical manufacturers pay a rebate equal to at least 23.1\% of the average price they earn on sales to retail pharmacies for brand-name drugs purchased by Medicaid beneficiaries. 
In this program, a manufacturer who wants its drug covered under Medicaid must enter into a rebate agreement with the Secretary of Health and Human Services and state that it will rebate a specified portion of the Medicaid payment for the drug to the states, which in turn share the rebates with the federal government.\footnote{Manufacturers must also enter agreements with other federal programs serving vulnerable populations. In exchange, Medicaid programs cover nearly all of the manufacturer's FDA-approved drugs, which are eligible for federal matching funds. Though the pharmacy benefit is a state option, all states cover it but administer pharmacy benefits in somewhat different ways within federal guidelines about pricing and rebates. For generic drugs, the rebate amount is 13\% of AMP, and there is no best price provision.} There are 1,982 brand names and 448 generic pharmaceutical manufacturing businesses in the US as of 2021.\footnote{IBISWorld-Industry Market Research, Reports, \& Statistics.} Approximately 600 drug manufacturers currently participate in this program (medicaid.gov).

Our simulation assumes that the mechanism works in several steps. First, the state determines the rebate portion. Next, manufacturers decide whether to participate or not. Lastly, after the entry decision is made, the manufacturers determine their retail prices. We assume that all manufacturers know each other's costs before the participation and pricing decisions, i.e., games of complete information. When all three manufacturers choose not to participate, the WIC consumers will purchase store brands.

The equilibrium strategies in the two-stage game of complete information are solved via backward induction. Given the configuration of the manufacturers' participation decisions and their marginal costs including the cost savings, we first solve for the optimal pricing game. We then can compute, for each configuration, all the manufacturers' profits. 

We then solve for the manufacturers' optimal participation decisions given their profits. We consider pure strategy Nash equilibrium. This entry game could admit multiple equilibria for some values of the cost vector and the predetermined rebates. In our simulation, we choose the rebate to be 55\% of the retail price with which each game admits a unique equilibrium. For rebates higher than 55\%, some markets have multiple equilibria, with which we do not have clear criteria to choose among all equilibria. To avoid this complication, we use rebates of 55\%. In some markets with a 55\% rebate, only two manufacturers choose to participate in the WIC program, but all three manufacturers participate in more than half of the markets. We then compute WIC and non-WIC consumers' surplus, manufacturer profit, and government expenditure.

 \subsection{Results}
\paragraph{Evaluating the WIC auction}
We first simulate the market outcomes to fully evaluate the impacts of the WIC procurement auction compared with the mechanism without auction. We present the comparison of market outcomes in Table \ref{Counterfactual1}. We find that WIC consumers are better off without the auction due to the availability of more options so that they can choose their preferred WIC brand, resulting in a higher consumer surplus. Moreover, fewer WIC consumers choose the outside option. In contrast, surprisingly, non-WIC consumers are worse off with the auction. This is because retail prices are lowered due to the spill-over effect and cost savings enjoyed by the auction winner. Also, fewer non-WIC consumers choose outside options 
with WIC auctions because they prefer products of the auction winner due to spill-over effects. The total consumer surplus decreased from 8.53 to 7.59 million dollars a month, around an 11\% reduction. 

Secondly, as expected, the government successfully contains the program cost using the auction mechanism. Specifically, the government expenditure decreases from 3.21 million dollars a month to 1.14 million dollars a month, a 65\% reduction in expenditure. 

Thirdly, procurement auction significantly reduces manufacturers' profits of Nestl\'{e}, the smallest manufacturer, implying that it faces the most significant negative impact. The profit of 
Nestl\'{e}, Abbott, and MJ drop by 75\%, 58\%, and 42\%, respectively in the current mechanism, compared with the one without competitive bidding.   

\paragraph{The Impact of Program Size}
We then investigate how WIC program size affects the market outcomes, mainly the manufacturers' profits, in the current auction procurement. We increase and decrease the number of WIC infants in the 2015 data by 10\%, then re-solve the equilibrium, and compute the market outcomes.  This simulation is very useful in evaluating the impact of the auction when the program size varies, a scenario in the late 1990s, when the number of WIC participants increases dramatically. Four manufacturers were active and participating in WIC auctions in the market when the government started implementing auctions to procure infant formulas. The smallest manufacturer, Wyeth, exited the market and the CEO announced that ``the increasing growth of the WIC Program contributed to the decision". 

We present the simulation results in Table \ref{CP2}, which demonstrates that the larger the program size, the more profits the two larger manufacturers have. By contrast, the profit of the smallest manufacturer Nestl\'{e}, which has a similar market share as Wyeth in the 1990s, decreases in the program size. This is consistent with Wyeth's claim that the increase in the number of WIC program participants pushes it out of the market. 

Additionally, it is expected that government expenditure increases in program sizes. Consumer surplus also increases in the program size and the changes are mainly from WIC consumers. The impact of program size on the surplus of non-WIC consumers is marginal, when the size increases by 10\%, the surplus increases only by 0.1\%.

\paragraph{Switching from auction to predetermined rebate} 
Last, we quantify the market outcomes if the government switches from the competitive bidding mechanism to the predetermined rebate and present the results in Table \ref{CP4}. 

We find that the government expenditure using the predetermined rebates is 27\% higher, a considerable increment. This is due to the fact that the 55\% rebate is relatively smaller than the rebate in the competitive bidding mechanism (with an average rebate being 85\%).\footnote{It is possible that the government can do better to contain cost by adopting the optimal predetermined rebate.} 

Non-WIC and WIC consumers are better and worse off, respectively, if predetermined rebates are adopted. Specifically, WIC consumers' surplus would be 26\% higher, which is a dramatic increase, while non-WIC consumers' surplus reduces by 5\%. The overall consumer surplus is 12.5\% higher in the mechanism with predetermined rebates.

The last and most interesting results are on manufacturers' profit. 
We find that three manufacturers are affected differently if the mechanism switches from competitive bidding to predetermined rebates. Nestl\'{e}, the smallest manufacturer among the three would benefit from the switching and enjoy a 47\% increase in profit. By contrast, both MJ and Abbott are harmed by the switching, with a 14.8\% and 30.3\% drop in profit, respectively. These results reinforce our finding in Section \ref{sec: estimation of supply side} that the procurement with auction serves as a barrier for small manufacturers to enter the market and to grow if they are already in the market.

To summarize, the mechanism with a predetermined rebate might be better than the auction in the sense that (1) the total consumer surplus increases while the government expenditure only rises slightly; (2) it protects manufacturers with small market shares and thus promotes competition in the market.

\section{Concluding remarks \label{sec:conclusions}}

We studied the impacts of the WIC program's procurement auction on manufacturers' behavior and consumers' welfare. We first observed  that manufacturers bid aggressively and discount their products substantially to the WIC program. Next, we introduced a structural model to capture manufacturers' competition in both non-WIC and WIC markets. By estimating the model using the NielsenIQ sales data and FNS' bidding data, we found that the winning manufacturers are operating below marginal costs in more than half of the WIC markets. They compensate their loss in the WIC markets by the gain in the non-WIC markets because a winner enjoys cost reductions and spill-over effects of demand. Based on the estimates of our structural model, we conducted Monte Carlo simulations to investigate the determinants of the market outcomes in the current mechanism of procurement auction and study the comparison between the current mechanism with the predetermined rebates adopted in MDRP. The most interesting finding is that the mechanism with a predetermined rebate could be better than the auction in the sense that the total consumer surplus increases while the government expenditure only rises slightly and it protects manufacturers with small market shares and thus promotes competition in the market. 

Although our method is developed in the context of infant formula, it might be extended to other social programs where government purchasing is involved. One caveat in our paper is that we do not model the manufacturers' bidding strategy directly. Instead, we use the reduced-form bidding function to capture the relationship between the bidding rebates and the market covariates. Such a reduced-form function is simple and useful for avoiding the complication of modeling the bidding strategy but also comes with a limitation of our analysis. That is, we cannot simulate the market outcomes by incorporating changes in the market structure, e.g., there is a new entrant in the industry. Modeling the manufacturer's bidding strategy in the WIC auction is challenging but exciting, which we leave for future research. 

Recently, Federal Trade Commission (FTC) investigates Abbott for possible collusion in WIC auctions.\footnote{Sandhu, R. (2023, May 24), Abbott, other formula makers face FTC investigation for collusion - WSJ. Retrieved from https://www.nasdaq.com/articles/abbott-other-formula-makers-face-ftc-investigation-for-collusion-wsj.} It will be interesting to explore possible anti-competitive conducts in the market and their impacts. 


\pagebreak

\paragraph{Disclaimer} Researcher(s)' own analyses calculated (or derived) based in part on data from Nielsen Consumer LLC and marketing databases provided through the NielsenIQ Datasets at the Kilts Center for Marketing Data Center at The University of Chicago Booth School of Business. The conclusions drawn from the NielsenIQ data are those of the researcher(s) and do not reflect the views of NielsenIQ. NielsenIQ is not responsible for, had no role in, and was not involved in analyzing and preparing the results reported herein. The findings and conclusions in this research have not been disseminated by the U.S Department of Agriculture and should not be construed to represent any agency determination or policy. All errors remain our own.
 Another research direction is to model pricing competition with two market segments where one is price insensitive but with price restriction. It will be important to study how government regulation connects to the pricing strategy. 
1. Predicted annually demand (volume) from 1998 to 2013 for each state and each manufacturer. 2. Average volume of formula an infant consumes in a year. 3. WIC demand evenly distributed for three manufacturers.

\bibliographystyle{aer}
\bibliography{overall}

\pagebreak

\textbf{{\LARGE {Appendix}}} \appendix

\numberwithin{equation}{section} \renewcommand\thesection{\Alph{section}}

\numberwithin{lemma}{section} \numberwithin{proposition}{section}

\section{Data appendix\label{Appendix: Data}}
This appendix further describes our data sources and manipulations
and tabulates the supplemental data used in our paper.\\

\noindent \textit{1. The number of infants.}
\begin{itemize}
  \item \textit{The total number of births.} The number of births by month and state from 1998-2018 is from the Centers for Disease Control and Prevention (CDC). The National Vital Statistics System, the Federal compilation of this data, is the result of the cooperation between the National Center for Health Statistics (NCHS) and the States to provide access to statistical information from birth certificates.
  \item \textit{The number of WIC infants.} The number of WIC participants from 1998-2013 is from the FNS of USDA, and 2014-18 is from WIC's website (coverage data). The coverage data cover monthly summaries of WIC participants for infants, children, and women by state.
\end{itemize}

\noindent\textit{2. Breastfeeding rates}
\begin{itemize}
  \item \textit{Overall breastfeeding rate.} The data on the overall breastfeeding rate, which covers non-WIC and WIC women, are from the National Immunization Survey (NIS) of the CDC. The data is  annual by state from 2000-2018. The breastfeeding rates include ever-breastfeeding at six months and 12 months. We use the rate of 12 months.

  \item \textit{Breastfeeding rate for WIC participants.} The annual breastfeeding rates are from the WIC Breastfeeding Data Local Agency Report of USDA starting from 2010 upon the requirement of the ``Healthy Hunger-Free Kids Act" of 2010. USDA classifies WIC participants as fully and partially breastfeeding. The rates are based on the population of WIC participants (2010-2018). \\
%
\end{itemize}

\noindent\textit{3. The ratio of WIC demand.} NielsenIQ's sales data are sampled, and it is not clear whether the sales are from WIC or non-WIC consumers in the data. We compute the WIC brand's non-WIC sales from its total monthly sales in a state as follows.
\begin{eqnarray*}
\hbox{non-WIC sales}=\left(1-\frac{(1-\hbox{WIC breastfeeding rate})*\hbox{\# of WIC infants}}{(1-\hbox{overall breastfeeding rate})*\hbox{\# of all infants}}\right)*\hbox{total sales}.
\end{eqnarray*}
Note that if an infant is eligible for WIC but does not participate, this infant is in the non-WIC group.\\

\noindent\textit{4. Demographics.}
\begin{itemize}
  \item \textit{Women labor participation rate.} Women labor participation rates by state from 1998 to 2018 are obtained from Expanded State Employment Status Demographic Data from Local Area Unemployment Statistics of US Bureau of Labor Statistics.\\

  \item \textit{High school eduction.} The data on the educational attainment of those above 25 years old (for both males and females) from 1998-2018 are obtained from the US Census Bureau, American Community Survey, 2010 1-Year Estimates, Table S. 1501. \\
  \item \textit{Median income.} The median household incomes by state from 1998-2018 are obtained from the U.S. Census Bureau. Source: U.S. Bureau of the Census, Current Population Survey, Annual Social and Economic Supplements. \\
  \item \textit{Race.} The population distribution of women's race/ethnicity from 1998-2018 is obtained from the Bridged-Race Population Estimates of the CDC. We consider percentages of Hispanic, White and non-Hispanic, Black and non-Hispanic, and Asian and non-Hispanic.

\end{itemize}

\noindent\textit{5. Cost shifters.} Cost shifters are variables that affect the costs of infant formula. We include raw milk prices, (industrial) electricity prices, retail wage, and distance between plants of manufacturers and markets (states) to capture the costs of raw materials, production, and transportation, respectively. Except for distance, all the cost shifters are monthly data by state.
\begin{itemize}
  \item \textit{Raw milk prices.} Monthly raw milk prices by state from 1998-2018 are collected from the National Agricultural Statistics Service of USDA. 
  

  \item \textit{Electricity prices.} Monthly industrial electricity prices from 1998-2018 are collected at the state level from US Energy Information Administration (EIA).
  
  
  \item \textit{Distance.} The distance between a manufacturer and a market (state) is defined as the average mileage between the plant(s) of the manufacturer and the three largest cities in the state.
\end{itemize}

 \section{Figures and Tables}

\begin{figure}[tbp]
\caption{Impacts of winning on prices}
\begin{multicols}{2}
    \includegraphics[width=\linewidth]{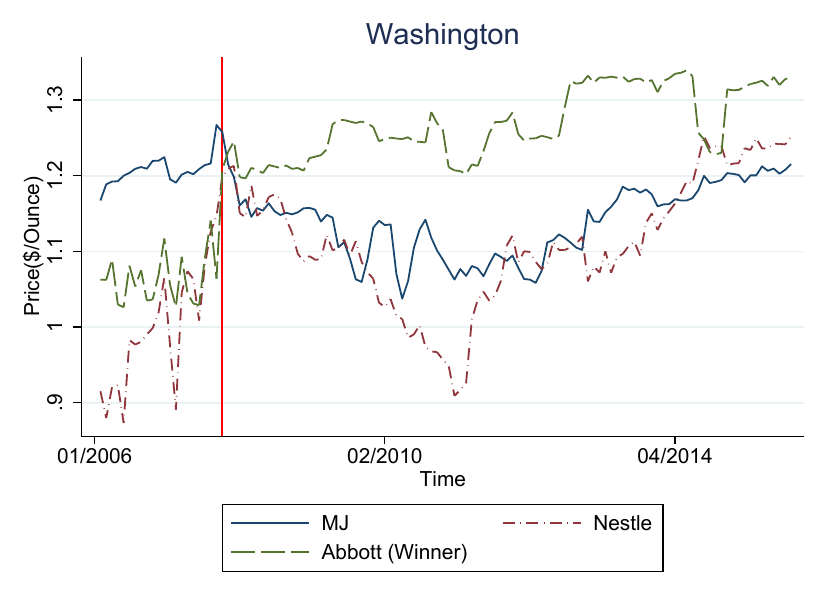}\par 
    \includegraphics[width=\linewidth]{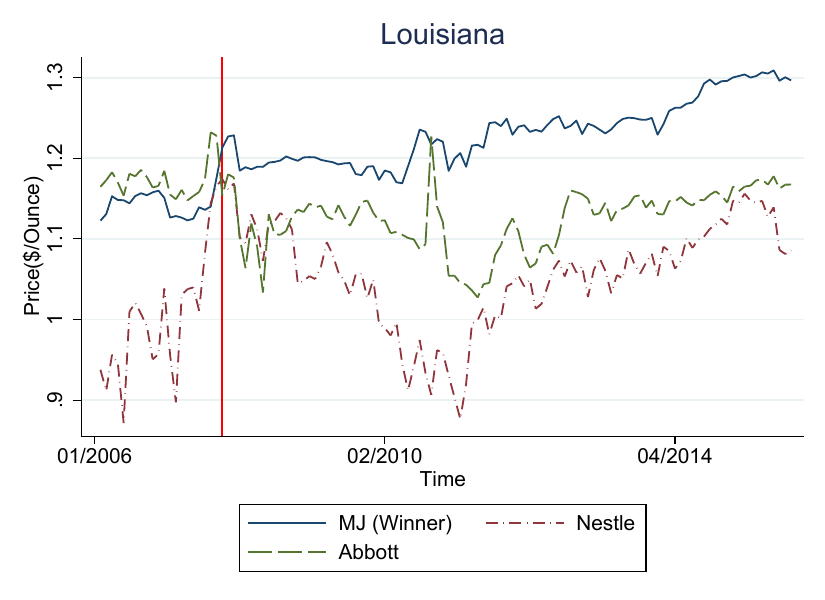}\par 
    \end{multicols}
\begin{multicols}{2}
    \includegraphics[width=\linewidth]{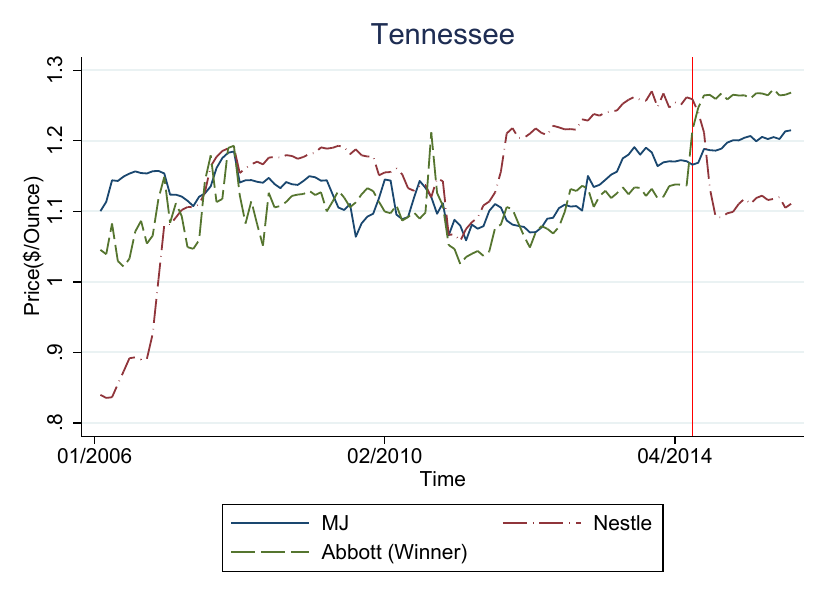}\par
    \includegraphics[width=\linewidth]{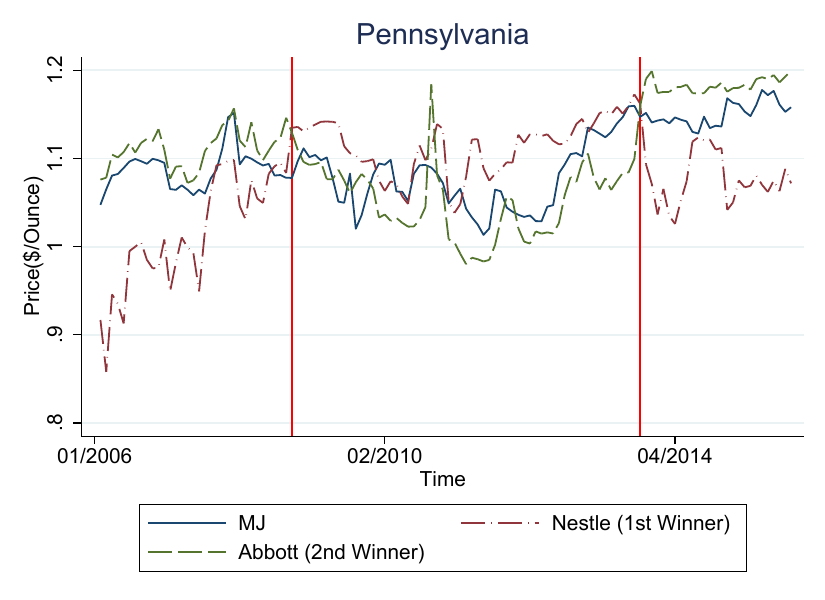}\par
\end{multicols}

\label{fig:impacts of winning on prices}\centering
\vspace{0.3cm}
\begin{minipage}{0.7\textwidth} 
{\footnotesize Note: The red vertical line indicates the starting time of a new WIC contract.
\par}
\end{minipage}
\end{figure}

 \begin{table}
\centering
\scalebox{0.8}{\begin{threeparttable}
\caption{Summary statistics of sales and prices}
\vspace{-0.6cm}
\begin{tabular}{lccccccccccc}
\multicolumn{11}{c}{}\\
\hline
\hline
 &\multicolumn{2}{c}{Full Sample} &	&\multicolumn{2}{c}{MJ}&	&\multicolumn{2}{c}{Abott}&&\multicolumn{2}{c}{Nestl\'{e}}		\\
	\cline{2-3}\cline{5-6}\cline{8-9}\cline{11-12}

	&	Mean  	&	Std.Dev	& &	Mean 	&	Std.Dev	&&	Mean  	&	Std.Dev&&	Mean  	&	Std.Dev\\
	
	\hline
Dollar Sales (\$1,000)	&	 444.43  &  741.83 	&&	626.58    &895.79&&	 564.19    &850.80	&&	 205.57   & 303.63	\\
Volume Sales (1,000 oz)	&	391.73 &    627.11	&&	532.56  &  764.82 &&	 470.67   & 707.43 	&&	 197.97  & 286.43  	\\
Unit Price (\$ per oz)	&	1.16 &    0.001	&&	1.21	&	0.45	&&	1.22	&0.24	&&	0.94	&	0.24	\\
\hline
\hline
\end{tabular}
\label{tab:SummaryStatistics}
\begin{tablenotes}
\item Note: The retail prices are deflated to 2015 dollars.
\end{tablenotes}
\end{threeparttable}}
\end{table}

 \begin{table}[htbp]
\centering
\begin{threeparttable}
\caption{Summary statistics of wholesale prices and rebates
\label{table: sumstat of whole sale and rebate}}
\begin{tabular}{l  c c c c c c  }\hline\hline
& \multicolumn{1}{c}{Variable} &\# of obs. & Mean & Std. Dev. & Min & Max  \\ \hline
 \multirow{2}{*}{Abbott}&wholesale & 78  & 1.051 & 0.059 & 0.976 & 1.180 \\
&rebate & 76  & 0.837 & 0.159 & 0.260 & 1.155 \\
 &ratio & 76  & 0.797 & 0.134 & 0.256 & 0.996 \\ \hline
 \multirow{2}{*}{MJ}&wholesale & 78  & 1.053 & 0.064 & 0.935 & 1.171 \\
&rebate & 73  & 0.865 & 0.137 & 0.403 & 1.140 \\
 &ratio  & 73  & 0.825 & 0.120 & 0.381 & 0.991 \\  \hline
 \multirow{2}{*}{Nestl\'{e}}&wholesale & 78  & 0.993 & 0.120 & 0.774 & 1.165 \\
&rebate & 51  & 0.880 & 0.147 & 0.482 & 1.180 \\
 &ratio & 51  & 0.864 & 0.085 & 0.613 & 1.052 \\  \hline

 \multirow{2}{*}{Overall}&wholesale & 234 & 1.032 & 0.090 & 0.774 & 1.180 \\
 &rebate & 200 & 0.858 & 0.149 & 0.260 & 1.180 \\
 &ratio & 200 & 0.824 & 0.121 & 0.256 & 1.052\\  \hline
\end{tabular}
 \begin{tablenotes}
\item \footnotesize{\textit{Source}: The Food and Nutrition Service of the United States Department of Agriculture.}
\item \footnotesize{\textit{Note}: Ratio is defined as the ratio of rebate over the wholesale price. The wholesale prices and rebates are in 2015 dollars per ounce.}
\end{tablenotes}
\end{threeparttable}
\end{table}

\begin{table}
\centering
\caption{Reduced-form evidence: Impact of winning on volume of sales}
\label{tab:results_sales}
\begin{center}
\scalebox{0.7}{\begin{threeparttable}
\begin{tabular}{lllllll} \hline\hline
 & (1) & (2) & (3) & (4) & (5) & (6) \\
\hline
log(Price) & -1.998** & -3.668*** & -2.689** & -1.143 & -4.196*** & -3.156** \\
 & (0.896) & (0.957) & (1.141) & (0.947) & (1.069) & (1.258) \\
Win & 0.807*** & 0.773*** & 0.754*** & 0.847*** & 1.123*** & 1.011*** \\
 & (0.036) & (0.038) & (0.038) & (0.084) & (0.086) & (0.083) \\
Auctioned brand & 3.026*** & 3.028*** & 2.967*** & 1.047*** & 1.132*** & 1.061*** \\
 & (0.041) & (0.040) & (0.053) & (0.063) & (0.057) & (0.053) \\
Win $\times$ auctioned brand & 0.402*** & 0.598*** & 0.508*** & -0.009 & 0.029 & -0.048 \\
 & (0.082) & (0.089) & (0.101) & (0.087) & (0.090) & (0.088) \\
Abbott &  &  &  & -1.688*** & -1.630*** & -1.676*** \\
 &  &  &  & (0.065) & (0.061) & (0.077) \\
Nestl\'{e} &  &  &  & -6.290*** & -6.298*** & -6.583*** \\
 &  &  &  & (0.112) & (0.105) & (0.105) \\
Win $\times$ Abbott &  &  &  & -0.803*** & -1.063*** & -0.952*** \\
 &  &  &  & (0.103) & (0.098) & (0.095) \\
Win $\times$ Nestl\'{e} &  &  &  & 0.061 & -0.327*** & -0.178* \\
 &  &  &  & (0.088) & (0.095) & (0.092) \\
Auctioned brand $\times$ Abbott &  &  &  & 1.758*** & 1.600*** & 1.633*** \\
 &  &  &  & (0.087) & (0.082) & (0.071) \\
Auctioned brand $\times$ Nestl\'{e} &  &  &  & 4.956*** & 4.785*** & 5.106*** \\
 &  &  &  & (0.155) & (0.152) & (0.160) \\
Win $\times$ auctioned brand $\times$ Abbott&  &  &  & 1.035*** & 1.331*** & 1.319*** \\
 &  &  &  & (0.148) & (0.145) & (0.167) \\
Win $\times$ auctioned brand $\times$ Nestl\'{e} &  &  &  & 0.691*** & 0.950*** & 0.928*** \\
 &  &  &  & (0.138) & (0.137) & (0.155) \\
Number of non-WIC infants & 0.208*** & 0.235*** & 0.157*** & 0.209*** & 0.246*** & 0.163*** \\
 & (0.003) & (0.017) & (0.021) & (0.003) & (0.017) & (0.021) \\
log(Median income) &  &  & 0.914*** &  &  & 1.127*** \\
 &  &  & (0.251) &  &  & (0.294) \\
Women labor participation &  &  & 0.068*** &  &  & 0.076*** \\
 &  &  & (0.009) &  &  & (0.010) \\
High school education &  &  & 0.023 &  &  & 0.025 \\
 &  &  & (0.014) &  &  & (0.015) \\
White &  &  & 4.182*** &  &  & 3.948*** \\
 &  &  & (1.260) &  &  & (1.184) \\
Constant & 7.986*** & 7.727*** & -9.958*** & 9.955*** & 9.769*** & -10.543** \\
 & (0.091) & (0.111) & (3.415) & (0.109) & (0.133) & (4.114) \\
 &  &  &  &  &  &  \\
 Product fixed effect & Yes & Yes & Yes & Yes & Yes & Yes \\
State fixed effect & No & Yes & Yes &No & Yes & Yes \\
Observations & 29,179 & 29,179 & 28,618 & 29,179 & 29,179 & 28,618 \\

\hline\hline
\end{tabular}

\begin{tablenotes}
\item \textit{Notes:} Dependent variable is the logarithm of sales volume. Win is a dummy that takes value one for all the brands of the winning manufacturer and zero otherwise. The auction brand is a dummy that takes value one for all the auctioned brands and zero otherwise.  The logarithm of price is instrumented by milk price, electricity rate, and the average distance between a manufacturer's production center to a market. 
\item Robust standard errors in parentheses. $***p<0.01$, $**p<0.05$, $*p<0.1$.
\end{tablenotes}
\end{threeparttable}
}
\end{center}
\end{table}

\begin{table}
\caption{Reduced-form evidence: Impact of winning on retail prices}
\label{tab:reduced_price}
\scalebox{0.7}{\begin{threeparttable}
\begin{tabular}{lllllll}
\hline
\hline
 & (1) & (2)  & (3) &(4)&(5) &(6) \\
 \hline
log(Volume) & -0.002** & 0.013*** & 0.012*** & -0.001 & 0.022*** & 0.021*** \\
 & (0.001) & (0.003) & (0.003) & (0.001) & (0.003) & (0.003) \\
Win & 0.017*** & 0.004 & 0.004 & 0.020*** & -0.008 & -0.007 \\
 & (0.002) & (0.003) & (0.003) & (0.005) & (0.005) & (0.005) \\
Auctioned brand & 0.023*** & -0.020* & -0.016 & 0.036*** & 0.012*** & 0.008* \\
 & (0.003) & (0.011) & (0.011) & (0.003) & (0.004) & (0.004) \\
Win $\times$ Auctioned brand & 0.082*** & 0.081*** & 0.081*** & 0.019*** & 0.020*** & 0.021*** \\
 & (0.002) & (0.003) & (0.003) & (0.005) & (0.004) & (0.004) \\
Abbott &  &  &  & 0.004 & 0.033*** & 0.026*** \\
 &  &  &  & (0.003) & (0.007) & (0.007) \\
Nestl\'{e} &  &  &  & 0.017** & 0.152*** & 0.147*** \\
 &  &  &  & (0.007) & (0.022) & (0.022) \\
Win $\times$ Abbott &  &  &  & -0.019*** & 0.019*** & 0.018*** \\
 &  &  &  & (0.005) & (0.005) & (0.005) \\
Win $\times$ Nestl\'{e} &  &  &  & -0.003 & -0.000 & -0.001 \\
 &  &  &  & (0.005) & (0.005) & (0.005) \\
Auctioned brand $\times$ Abbott &  &  &  & -0.061*** & -0.102*** & -0.094*** \\
 &  &  &  & (0.003) & (0.008) & (0.007) \\
Auctioned brand $\times$ Nestl\'{e} &  &  &  & -0.111*** & -0.219*** & -0.215*** \\
 &  &  &  & (0.007) & (0.018) & (0.018) \\
Win $\times$ auctioned brand $\times$ Abbott &  &  &  & 0.106*** & 0.086*** & 0.085*** \\
 &  &  &  & (0.005) & (0.005) & (0.005) \\
Win $\times$ auctioned brand $\times$ Nestl\'{e}&  &  &  & 0.096*** & 0.080*** & 0.079*** \\
 &  &  &  & (0.006) & (0.006) & (0.006) \\
Average distance &  &  & -0.002 &  &  & 0.003 \\
 &  &  & (0.004) &  &  & (0.004) \\
Electricity rate &  &  & 0.043 &  &  & -0.038 \\
 &  &  & (0.044) &  &  & (0.045) \\
Raw milk price &  &  & 0.019*** &  &  & 0.019*** \\
 &  &  & (0.002) &  &  & (0.002) \\
Constant & 0.105*** & -0.001 & -0.029 & 0.105*** & -0.120*** & -0.139*** \\
 & (0.007) & (0.027) & (0.028) & (0.009) & (0.034) & (0.034) \\
 &  &  &  &  &  &  \\
Product fixed effect & Yes & Yes & Yes & Yes & Yes & Yes \\
State fixed effect & No & Yes & Yes &No & Yes & Yes \\
Observations & 28,632 & 28,632 & 28,618 & 28,632 & 28,632 & 28,618 \\
 
 \\
\hline\hline
\end{tabular}
\begin{tablenotes}
\item \textit{Notes:} Dependent variable: logarithm of price (dollar per ounce). Win is a dummy that takes value one for all the brands of the winning manufacturer and zero otherwise. The auction brand is a dummy that takes value one for all the auctioned brands and zero otherwise. The logarithm of volume is instrumented by the number of non-WIC infants, the logarithm of median state income, and the women's labor participation rate.
\item Robust standard errors in parentheses. $***p<0.01$, $**p<0.05$, $*p<0.1$.
\end{tablenotes}
\end{threeparttable}}
\end{table}

\begin{table}
 \caption{Results of demand estimation}
\label{table: demand}
\scalebox{0.8}{\begin{threeparttable}
\begin{tabular}{llll} \hline
             & Mean    & Std & Income \\ \hline
Price        & -2.3818*** & 0.0196            & 0.0190*** \\
             & (0.0662)  & (1.0724)        & (0.0053) \\
Win      & 0.3160***  &                   &        \\
             & (0.0220)  &                   &        \\
Auctioned brand & 3.1773***  &                   &        \\
             & (0.0511)  &                   &        \\
WIC brand    & 0.9585***  &                   &        \\
             & (0.0367)  &                   &        \\
Spit up        & -2.8360*** &                   &        \\
             & (0.2273)  &                   &        \\
Prebiotics   & -0.9545*** &                   &        \\
             & (0.1159)  &                   &        \\
Nestl\'{e}       & -2.4457*** &                   &        \\
             & (0.1147)  &                   &        \\
Abbott       & -0.5912*** &                   &        \\
             & (0.0448)  &                   &       \\ 
Constant     & 0.0554  & 0.0436            &        \\
             & (0.1377)  & (1.4057)            &        \\
Observations & 36,224\\
            \hline\hline
\end{tabular}
\begin{tablenotes}
\item \textit{Notes:} Standard errors are in parentheses. $***p<0.01$, $**p<0.05$, $*p<0.1$. 
 \end{tablenotes}
\end{threeparttable}
}
 \end{table}

\begin{table}
\begin{center}
\caption{Median own- and cross-elasticities}
\label{table: elasticities} 
\scalebox{0.7}{\begin{threeparttable}
\begin{tabular}{llcccccccc} \hline \hline
                        &                 & \multicolumn{2}{c}{MJ} &        & \multicolumn{2}{c}{Nestl\'{e}} &    & \multicolumn{2}{c}{Abbott}     \\ 
 \cline{3-4}   \cline{6-7} \cline{9-10}                    
Manufacturer                   & Product         & Auction & non-Auction && Auction & non-Auction && Auction & non-Auction \\ \hline
\multirow{2}{*}{MJ}     & Auction    & -2.2575      & 0.4559 &         & 0.4690       & 0.4559 &         & 0.4702       & 0.4618          \\
                        & non-Auction & 0.0252       & -2.7414 &        & 0.0250       & 0.0250          & &0.0251       & 0.0158          \\ \hline
\multirow{2}{*}{Nestl\'{e}} & Auction    & 0.1080       & 0.1086 &         & -2.3936      & 0.1086        &  & 0.1078       & 0.1101          \\
                        & non-Auction & 0.0075       & 0.0040 &         & 0.0075       & -2.5028         & &0.0075       & 0.0004          \\ \hline
\multirow{2}{*}{Abbott} & Auction    & 0.7029       & 0.6678 &         & 0.7011       & 0.6678        &  & -1.9801      & 0.6759          \\
                        & non-Auction & 0.0245       & 0.0112  &        & 0.0245       & 0.0112          & &0.0245       & -2.6040        \\
                        \hline \hline
\end{tabular}
\end{threeparttable}}
\end{center}
\end{table}

\begin{table}[]
\caption{Cost functions and WIC pricing adjustments}
\label{pricing1}
\scalebox{0.8}{\begin{threeparttable}
\begin{tabular}{llcc}             \hline \hline
            &                                      & Estimate     & Std. \\
            \hline
            & Nestl\'{e}                               & 0.0307  & 0.0124         \\
            & Abbott                               & -0.0541 & 0.0050         \\
            & auctionBrand                         & -0.0255 & 0.0055         \\
            & win                                  & -0.1465 & 0.0028         \\
            & win*Nestl\'{e}                           & 0.0555  & 0.0039         \\
            & win*Abbott                           & -0.0341 & 0.0074         \\
            & win*auctionBrand                     & 0.0077  & 0.0049         \\
            & win*auctionBrand*Nestl\'{e}              & -0.0838 & 0.0072         \\
    cost    & win*auctionBrand*Abbott              & 0.0122  & 0.0092         \\
            & spitup                               & -0.0728 & 0.0247         \\
            & prebiotics                           & 0.1263  & 0.0125         \\
            & electricity price                     & 0.4212  & 0.1397         \\
            & raw milk                             & -0.0262 & 0.0167         \\
            & distance                             & -0.0385 & 0.0065         \\
            & electricity price*electricity price & -3.5835 & 0.7187         \\
            & raw milk*raw milk                  & 0.0198  & 0.0055         \\
            & distance*distance                  & 0.0262  & 0.0031         \\  
           & constant                             & 0.5111  & 0.0185         \\ \hline
 
            & MJ                                   & 0.01    &                \\
pricing adjustments& Nestl\'{e}                               & 0.11    &                \\
            & Abbott                               & 0.07    &               \\
            \hline \hline
\end{tabular}
\end{threeparttable}}
\end{table}

\begin{table}[]
\caption{Summary statistics of the estimated marginal costs and markups }
\label{est: markup}
\scalebox{0.8}{\begin{threeparttable}
\begin{tabular}{lllllllll} \hline \hline
                               &              & \multicolumn{3}{c}{Losing} & &\multicolumn{3}{c}{Winning} \\
                               \cline{3-5}\cline{7-9}
                               &              & median   & mean   & std    && median   & mean    & std    \\ \hline
\multirow{4}{*}{marginal cost} & overall      & 0.594    & 0.587  & 0.133  & &0.437    & 0.440   & 0.160  \\
                               & MJ & 0.636    & 0.634  & 0.125  & &0.485    & 0.480   & 0.158  \\
                               & Nestl\'{e}       & 0.569    & 0.557  & 0.137  & &0.423    & 0.432   & 0.161  \\
                               & Abbott       & 0.579    & 0.566  & 0.081  & &0.403    & 0.393   & 0.145  \\
                               \hline
\multirow{4}{*}{markup}        & overall      & 0.451    & 0.464  & 0.085  & &0.611    & 0.622   & 0.128  \\
                               & MJ & 0.453    & 0.464  & 0.091  & &0.595    & 0.605   & 0.131  \\
                               & Nestl\'{e}       & 0.445    & 0.461  & 0.086  & &0.593    & 0.606   & 0.124  \\
                               & Abbott       & 0.465    & 0.478  & 0.057  & &0.659    & 0.670   & 0.118 \\
                               \hline \hline
\end{tabular}
\begin{tablenotes}
\item \textit{Notes:} The results are calculated using the marginal costs and price adjustments in Table \ref{pricing1}. The calculation of markups upon winning does not take into account rebates. 
 \end{tablenotes}
\end{threeparttable}
}
\end{table}

\begin{table}
\begin{center}
\caption{Manufacturers' profits }
\label{table: elasticities} 
\scalebox{0.7}{\begin{threeparttable}
\begin{tabular}{lcccccccc} \hline \hline
     & \multicolumn{2}{c}{MJ} &        & \multicolumn{2}{c}{Nestl\'{e}} &    & \multicolumn{2}{c}{Abbott}     \\ 
        \cline{2-3}                       \cline{5-6}                           \cline{8-9}                    
               Profit   &          Mean & Median && Mean & Median && Mean & Median\\ \hline
All& 0.464 & 0.087 &  & 0.264 & 0.054 &  & 1.063 & 0.464 \\
non-WIC& 1.508 & 0.920 &  & 0.745 & 0.584 &  & 1.328 & 0.678 \\
WIC & -1.044 & -0.568 &  & -0.482 & -0.295 &  & -0.265 & -0.134 \\  
                        \hline \hline
\end{tabular}
\begin{tablenotes}
\item \textit{Notes:} All values are in millions of 2015 dollars. 
\end{tablenotes}
\end{threeparttable}}
\end{center}
\end{table}

\begin{table}[]
\caption{Approximation of the rebate bidding strategies}
\scalebox{0.8}{\begin{threeparttable}
\begin{tabular}{lllllll} \hline \hline
                    & (1) &(2) & (3) & (4) & (5) &(6) \\ \hline
log(wholesale price)     & 1.208***    & 1.409***    & 1.575***    & 1.544***    & 1.561***    & 2.143***    \\
                    & (0.13)      & (0.26)      & (0.30)      & (0.27)      & (0.27)      & (0.34)      \\
log(rival wholesale price)      &             & -0.245      & -0.438      & -0.683*     & -0.622*     & -1.129**    \\
                    &             & (0.24)      & (0.30)      & (0.29)      & (0.31)      & (0.39)      \\
log(distance)       &             &             & 0.046      & 0.035      & 0.026      & 0.048      \\
                    &             &             & (0.05)      & (0.05)      & (0.04)      & (0.05)      \\
log(rival distance)    &             &             & -0.030     & -0.064     & -0.081     & -0.103      \\
                    &             &             & (0.06)      & (0.06)      & (0.06)      & (0.07)      \\
log(income)         &             &             &             & -0.313*     & -0.398*     & -0.327      \\
                    &             &             &             & (0.13)      & (0.17)      & (0.17)      \\
log($\#$ of WIC infants)      &             &             &             & -0.134**    & -0.150***   & -0.126**    \\
                    &             &             &             & (0.04)      & (0.04)      & (0.04)      \\
log($\#$ of non-WIC infants)   &             &             &             & 0.171***    & 0.205***    & 0.182***    \\
                    &             &             &             & (0.05)      & (0.05)      & (0.05)      \\
log(raw milk price)       &             &             &             &             & -0.186      & -0.211*     \\
                    &             &             &             &             & (0.11)      & (0.11)      \\
log(electricity price)   &             &             &             &             & 0.022       & 0.023      \\
                    &             &             &             &             & (0.09)      & (0.08)      \\
log(contract length) &             &             &             &             & -0.062     & -0.066*    \\
                    &             &             &             &             & (0.03)      & (0.03)      \\
Nestl\'{e}              &             &             &             &             &             & 0.090*     \\
                    &             &             &             &             &             & (0.04)      \\
Abbott              &             &             &             &             &             & -0.035     \\
                    &             &             &             &             &             & (0.04)      \\
constant & -0.215*** & -0.214*** & -0.208*** & 0.940   & 1.831  & 1.597  \\
       & (0.01)    & (0.01)    & (0.02)    & (0.49) & (0.98) & (0.96) \\
Observations                   & 200         & 200         & 200         & 200         & 200         & 200         \\
adjusted $R^2$           & 0.211       & 0.209       & 0.204       & 0.267       & 0.282       & 0.327      \\
\hline\hline
\end{tabular}
\begin{tablenotes}
\item \textit{Notes:} Dependent variable: the logarithm of rebate.
\item Standard errors are in parentheses. $***p<0.01$, $**p<0.05$, $*p<0.1$. 
 \end{tablenotes}
\end{threeparttable}} 
\label{table: rebate}
\end{table}

\begin{table}[htbp] \centering
\caption{Counterfactual analysis: Impact of WIC auction}
\label{Counterfactual1}
\scalebox{0.75}{\begin{threeparttable}
\vspace{-0.3cm}
\begin{tabular}{ccccc} \hline \hline
                       & no Auction & Auction & diff & pct. diff. \\ \hline
$GEXP$  & 3.210      & 1.140        & -2.070     & -64.5\%    \\
$TS$                     & 11.373     & 8.920        & -2.452     & -21.6\%    \\
$CS$                    & 8.531      & 7.585        & -0.946     & -11.1\%    \\
$\pi_{all}$                & 2.841      & 1.335        & -1.507     & -53.0\%    \\
$CS_{WIC}$                 & 5.432      & 4.318        & -1.114     & -20.5\%    \\
$CS_{non-WIC}$            & 3.099      & 3.267        & 0.168      & 5.4\%      \\
$\pi_{MJ}$                & 1.198      & 0.498        & -0.700     & -58.4\%    \\
$\pi_{Nestl\acute{e}}$            & 0.356      & 0.088        & -0.268     & -75.3\%    \\
$\pi_{Abbott}$            & 1.287      & 0.748        & -0.538     & -41.8\%   \\
\hline \hline
\end{tabular}
\begin{tablenotes}
\item \textit{Notes:} All values are in millions of 2015 dollars. $GEXP$ is government expenditure on WIC,
$CS$ is consumer surplus, $TS$ is total surplus, and $\pi$ is profit.
\end{tablenotes}
\end{threeparttable}
}
\end{table}

\begin{table}[htbp] \centering
\caption{Counterfactual analysis: The impact of WIC program size}
\label{CP2}
\vspace{-0.3cm}
\scalebox{0.75}{
\begin{threeparttable}
\begin{tabular}{cccccccccccc} \hline \hline
                     &  & \multicolumn{3}{c}{WIC size decreases by 10\%} && \multicolumn{3}{c}{WIC size increases by 10\%} \\

                      \cline{3-5}\cline{7-9}
                     &  data    & value        & diff         & pct. diff.      &  & value         & diff        & pct. diff.        \\
                     \hline
$GEXP$       & 1.140 & 0.993         & -0.148        & -12.9\%         & & 1.285         & 0.145         & 12.7\%           \\
$TS$         & 8.920 & 8.419         & -0.501        & -5.6\%         &  & 9.399         & 0.479         & 5.4\%            \\
$CS$        & 7.585 & 7.119         & -0.467        & -6.2\%         &  & 8.036         & 0.451         & 5.9\%            \\
$\pi_{all}$     & 1.335 & 1.300         & -0.035        & -2.6\%        &   & 1.363         & 0.028         & 2.1\%            \\
$CS_{WIC}$   & 4.318 & 3.857         & -0.461        & -10.7\%       &   & 4.766         & 0.448         & 10.4\%           \\
$CS_{non-WIC}$   & 3.267 & 3.262         & -0.006        & -0.2\%        &   & 3.270         & 0.003         & 0.1\%            \\
$\pi_{MJ}$      & 0.498 & 0.481         & -0.017        & -3.5\%       &    & 0.509         & 0.011         & 2.2\%            \\
$\pi_{Nestl\acute{e}}$ & 0.088 & 0.092         & 0.004         & 4.0\%        &    & 0.085         & -0.003        & -3.3\%           \\
$\pi_{Abbott}$ & 0.748 & 0.728         & -0.021        & -2.8\%        &   & 0.768         & 0.020         & 2.6\%             \\
 \hline \hline
\end{tabular} 
\begin{tablenotes}
\item \textit{Notes:} All values are in millions of 2015 dollars. $GEXP$ is government expenditure on WIC,
$CS$ is consumer surplus, $TS$ is total surplus, and $\pi$ is profit.
\end{tablenotes}
\end{threeparttable}
}
\end{table}

\begin{table}[htbp]
\caption{Counterfactual analysis: auction versus pre-determined rebate}
\vspace{-0.3cm}
\label{CP4}
\scalebox{0.75}{
\begin{threeparttable}
\begin{tabular}{ccccc} \hline \hline
           & Auction & rebate=55\% & diff  & pct. diff. \\ \hline
$GEXP$      & 1.140        & 1.445                 & 0.304  & 26.7\%     \\
$TS$          & 8.920        & 9.607                 & 0.686  & 7.7\%      \\
$CS$          & 7.585        & 8.530                 & 0.945  & 12.5\%     \\
$\pi_{all}$     & 1.335        & 1.076                 & -0.259 & -19.4\%    \\
$CS_{WIC}$     & 4.318        & 5.431                 & 1.113  & 25.8\%     \\
$CS_{non-WIC}$ & 3.267        & 3.099                 & -0.168 & -5.1\%     \\
$\pi_{MJ}$    & 0.498        & 0.424                 & -0.074 & -14.8\%    \\
$\pi_{Nestl\acute{e}}$& 0.088        & 0.130                 & 0.042  & 47.3\%     \\
$\pi_{Abbott}$ & 0.748        & 0.522                 & -0.227 & -30.3\%     \\
\hline \hline
\end{tabular}
\begin{tablenotes}
\item \textit{Notes:} All values are in millions of 2015 dollars. We set the pre-determined rebate to be 55\%. $GEXP$ is government expenditure on WIC, $CS$ is consumer surplus, $TS$ is total surplus, and $\pi$ is profit.
\end{tablenotes}
\end{threeparttable}
}
\end{table}

\end{document}